# Dissociable changes in functional network topology underlie early category learning and development of automaticity

Fabian A. Soto[1], Danielle S. Bassett[2,3], & F. Gregory Ashby[1]

[1] Department of Psychological & Brain Sciences, University of California, Santa Barbara, CA 93106 USA
[2] Department of Bioengineering, University of Pennsylvania, Philadelphia, PA 19104 USA
[3] Department of Electrical & Systems Engineering, University of Pennsylvania, Philadelphia, PA 19104 USA

**Correspondence to:** Fabian A. Soto. Department of Psychological & Brain Sciences,
University of California, Santa Barbara. Santa Barbara, CA 93106, USA. E-mail: fabian.soto@psych.ucsb.edu

## Abstract

Recent work has shown that multimodal association areas–including frontal, temporal and parietal cortex–are focal points of functional network reconfiguration during human learning and performance of cognitive tasks. On the other hand, neurocomputational theories of category learning suggest that the basal ganglia and related subcortical structures are focal points of functional network reconfiguration during early learning of some categorization tasks, but become less so with the development of automatic categorization performance. Using a combination of network science and multilevel regression, we explore how changes in the connectivity of small brain regions can predict behavioral changes during training in a visual categorization task. We find that initial category learning, as indexed by changes in accuracy, is predicted by increasingly efficient integrative processing in subcortical areas, with higher functional specialization, more efficient integration across modules, but a lower cost in terms of redundancy of information processing. The development of automaticity, as indexed by changes in the speed of correct responses, was predicted by lower clustering (particularly in subcortical areas), higher strength (highest in cortical areas) and higher betweenness centrality. By combining neurocomputational theories and network scientific methods, these results synthesize the dissociative roles of multimodal association areas and subcortical structures in the development of automaticity during category learning.

## Introduction

Network science provides a set robust tools that are increasingly used to describe and understand neural systems (Bullmore and Sporns, 2009; Sporns, 2014). Neurons or brain regions are represented as network nodes, and structural or functional connections between regions are represented as network edges. Recent studies demonstrate that the topology of functional brain networks can reconfigure quickly as the result of learning (Bassett et al., 2011, 2013b) and task engagement (Bassett et al., 2006; Kitzbichler et al., 2011; Ekman et al., 2012; Fornito et al., 2012). In several cases, this reconfiguration leads to more integrated and less segregated processing (Kitzbichler et al., 2011; Ekman et al., 2012; Cole et al., 2014), and involves strong

This work was supported in part by Grant No. W911NF-07-1-0072 from the U.S. Army Research Office through the Institute for Collaborative Biotechnologies. DSB was supported by the Alfred P. Sloan Foundation, the Army Research Laboratory through contract no. W911NF-10-2-0022 from the U.S. Army Research Office, and awards #BCS-1441502 and #BCS-1430087 from the National Science Foundation.

reconfiguration in some nodes, while global network properties can remain relatively stable (Bassett et al., 2006; Moussa et al., 2011; Rzucidlo et al., 2013). In particular, nodes in multimodal association areas–within frontal, temporal and parietal cortex–flexibly change their community affiliation during learning (Bassett et al., 2011, 2013b), their connectivity pattern during rule application and preparatory attention (Ekman et al., 2012; Cole et al., 2013), and the cost-efficiency of their connectivity during accurate performance of working memory tasks (Bassett et al., 2009).

Despite these results, it is unlikely that connectivity changes in cortical association areas underlie functional network reconfigurations across all tasks. For example, connectivity changes and integrative processing in the basal ganglia are likely to be of upmost importance during initial learning of some categorization tasks (Ashby and Ennis, 2006). With the development of automatic performance, the basal ganglia might lose their central role (Waldschmidt and Ashby, 2011) as direct sensory-motor connections are formed (DeGutis and D'Esposito, 2009). This implies a switch to a more direct, efficient form of integration that does not require mediation by multiple synaptic connections (see Ashby et al., 2007).

These two stages of learning–category learning and automaticity–should be reflected differently in different behavioral measures of performance (Ashby et al., 2007). Relatively fast changes in accuracy tend to reflect initial learning supported by the basal ganglia and related subcortical areas, whereas slower changes in the speed of correct responses tend to reflect the switch to cortically controlled automatic performance. This dissociation in behavioral measures can be used to study whether and how changes in functional networks are related to different stages of category learning. We can expect changes in the connectivity of subcortical areas–instead of cortical association areas–to predict initial category learning best. Furthermore, this central role of the basal ganglia should be more apparent in the prediction of accuracy than in the prediction of response times.

Here we explore these predictions using a combination of network science and multilevel regression (Gelman and Hill, 2007). We study how changes in the connectivity of small brain regions can predict behavioral changes during extensive training in a task known to foster procedural category learning (Ashby et al., 2003).

## Materials and Methods

## Experimental procedures

*Participants.* Ten healthy undergraduate students from the University of California, Santa Barbara (6 males, 4 females) voluntarily participated in this study in exchange for course credit or a monetary compensation. This is a small, but sufficient sample size (Snijders and Bosker, 2012) that has been shown to provide unbiased estimates of regression coefficients in multilevel regression (Bell et al., 2014; Maas and Hox, 2005; see discussion in the supplementary material). All participants gave their written informed consent to participate in the study. The institutional review board of the University of California, Santa Barbara, approved all procedures in the study.

Standard univariate and multivariate analyses of the imaging data acquired on this sample have been previously reported (Waldschmidt and Ashby, 2011; Soto et al., 2013). We excluded one person from the full sample of eleven participants due to incomplete data.

*Behavioral task.* The stimuli were circular sine-wave gratings of constant contrast and size (see example in Figure 1A) that varied in orientation from 20° to 110° and in frequency from 0.25 to 3.58 cycles per stimulus width. Figure 1B shows the category structure used to train participants; each dot in the figure represents a different stimulus and the dotted line represents the boundary separating the two categories. Previous research suggests that this task is mastered by people through procedural learning (e.g., Ashby et al., 2003; Maddox et al., 2004). During each trial, participants were presented with one of these stimuli and had to identify the category to which the stimulus belonged by pressing a button; this was followed by feedback indicating the accuracy of the response. Stimuli were presented and responses were recorded using MATLAB augmented with the Psychophysics Toolbox (Brainard, 1997), running on a Macintosh computer. For a more detailed description of the stimuli and apparatus, see (Helie et al., 2010).

The experiment consisted of 23 sessions of training in the categorization task, four of which were conducted in the MRI scanner. The training sessions were carried out over 23 consecutive workdays, one session per day. The scanning sessions were sessions 2, 4, 10 and 20, and each consisted of 6 blocks of 80 stimuli, for a total of 480 stimuli per session. Participants selected their responses through response boxes, where the button box in their left hand was correct for the category at the top-left of the bound in Figure 1B, and the button box in their right hand was correct for the category at the bottom-right of the bound in Figure 1B. Feedback was displayed for 2 s and consisted of a green check mark for correct responses or a red "X" mark for incorrect responses. If it took more than 2 s for the participant to respond, a black dot was displayed indicating that the response was too slow. Half of the trials included the presentation of a cross-hair before the stimulus presentation.

The 19 sessions of categorization training outside the scanner were similar to the scanner session, but carried out on a Macintosh computer. For a more detailed description of these sessions, see (Helie et al., 2010).

*Neuroimaging.* A rapid event-related fMRI procedure was used. Images were obtained using a 3T Siemens TIM Trio MRI scanner at the University of California, Santa Barbara Brain Imaging Center. The scanner was equipped with an 8-channel phased array head coil. Cushions were placed around the head to minimize head motion. A localizer, a GRE field mapping (3 mm thick; FOV: 192 mm; voxel: 3×3×3 mm; FA=60°), and a T1-flash (TR=15 ms; TE=4.2 ms; FA=20°; 192 sagittal slices 3-D acquisition; 0.89 mm thick; FOV: 220 mm; voxel: 0.9×0.9×0.9 mm; 256×256 matrix) were obtained at the beginning of each scanning session, and an additional GRE field-mapping scan was acquired at the end of each scanning session. Functional runs used a T2*-weighted single shot gradient echo, echo-planar sequence sensitive to BOLD contrast (TR: 2000 ms; TE: 30 ms; FA: 90°; FOV: 192 mm; voxel: 3×3×3 mm) with generalized auto calibrating partially parallel acquisitions (GRAPPA). Each scanning session lasted approximately 90 min.

## Data analysis

*fMRI data preprocessing.* The data series from each block was preprocessed using FEAT (fMRI Expert Analysis Tool) version 5.98 in FSL (www.fmrib.ox.ac.uk/fsl). Preprocessing included motion-correction to the middle volume in the series using tri-linear interpolation with six degrees of freedom in MCFLIRT (Jenkinson et al., 2002), slice timing correction (via Fourier time-series phase-shifting), BET brain extraction, and a high pass filter with a cutoff of 50 s. The data were not spatially smoothed during preprocessing to avoid artificially increasing the correlation between adjacent network nodes (Achard et al., 2006). Each functional scan was registered to the corresponding structural scan using FLIRT (Jenkinson and Smith, 2001;

Jenkinson et al., 2002) linear registration with its default settings. Each structural scan was registered to the MNI152-T1-2 mm standard brain using FNIRT (Andersson et al., 2007) nonlinear registration with its default settings. The resulting linear and nonlinear transformations were jointly used to transform the data series from subject space to standard space using tri-linear interpolation.

*Definition of regions of interest.* We were interested in measuring the functional connectivity of local clusters of voxels within regions of interest (ROIs) that are thought to be part of the brain network involved in category learning. With this goal in mind, we started by defining 36 anatomical ROIs previously reported to be involved in visual category learning and automaticity (for a more in-depth discussion of the hypothesized functional role of each ROI in categorization and the empirical evidence, see Soto et al., 2013).

The anatomical boundaries of each ROI were created in MNI152-T1-2 mm standard space using atlases included in FSL (Harvard-Oxford structural atlases, Oxford thalamic connectivity probability atlas, and Juelich histological atlas). Visual areas included the extrastriate cortex and inferotemporal cortex. Motor areas included primary motor cortex and premotor cortex. The premotor cortex was divided into the supplementary motor area (SMA), pre-SMA, ventral premotor area and dorsal premotor area as defined by (Picard and Strick, 2001). Subcortical areas included the caudate, putamen, pallidum, and thalamic areas. The caudate was divided into a head region and a body and tail region according to (Nolte, 2008). Using the Oxford thalamic connectivity probability atlas, we defined the medial dorsal nucleus of the thalamus as the thalamic area connected to prefrontal cortex, and the ventral anterior and ventral lateral nuclei of the thalamus as the thalamic area connected to primary motor and premotor cortices (Martin, 2003). Other cortical areas included the anterior cingulate cortex (ACC), prefrontal cortex (PFC) and hippocampus. The medial and posterior parts of the ACC were extracted following (Vogt et al., 2004). Following the definition given by (Petrides and Pandya, 2004), the dorsolateral PFC was extracted by joining the superior and middle frontal gyri and subtracting all premotor areas. Ventrolateral PFC was extracted using an inclusive definition from (Petrides and Pandya, 2004), which includes Brodmann areas 44, 45 and 47. For each of the aforementioned areas, we defined a left and a right ROI.

*Definition of network nodes.* The study of functional networks requires a definition of nodes and a measure of the functional connectivity between pairs of such nodes. Such functional connectivity measures are stored in an adjacency matrix **W**, with cell $W_{ij}$ representing the functional connectivity between nodes $i$ and $j$. The matrix **W** is the starting point for all network-based analyses (see below). Two common ways to define nodes in the study of functional networks are (i) the areas obtained from a brain atlas and (ii) spheres of a given radius centered at coordinates of interest (Varoquaux and Craddock, 2013). Here we take an intermediate strategy between those two approaches. We focused on the atlas-based ROIs defined in the previous section, which are thought to cover most of the brain network involved in visual category learning (see Soto et al., 2013). However, averaging the signal from such large areas could make it difficult to determine whether small clusters of voxels within each area are specifically related to categorization performance. Therefore, we subdivided each ROI into small clusters of voxels. Starting with a brain mask in MNI152-T1-2 mm standard space, we defined a rectangular box that enclosed this mask and divided it into 12×12×12 mm cubes (6×6×6 voxels). This size is comparable to the spheres of 5-10 mm radius commonly used in the literature (Power et al., 2011; Varoquaux and Craddock, 2013). If at least 75 voxels (~35%) of a cube fell inside an ROI, then the overlap between the cube and the ROI was defined as a node (for a similar approach, see Meunier et al., 2009). There was no spatial overlap between nodes (i.e., each voxel was included in only one node). This resulted in a total of 742 node masks, an order of magnitude larger than the common Automated Anatomical Labeling (AAL) atlas (Achard et al., 2006), finely covering the visual category learning network.

*Computation of functional connectivity matrices.* We built a functional connectivity matrix from each of the 24 pre-processed data series (from 6 separate blocks in each of 4 scanning sessions) separately for every participant. We chose to focus on task-related functional connectivity, and therefore removed all volumes before the first stimulus presentation and 20 seconds after the last stimulus presentation. The time series of each voxel in a node mask were averaged and the maximum-overlap discrete Daubechies 4 wavelet transform (Percival and Walden, 2000) of this averaged signal was computed using the WMTSA toolbox for MATLAB version 0.2.6 (Cornish, 2006). The wavelet transform was computed at three scales and, because the sampling frequency (TR) was 2 s, scale one corresponded approximately to 0.125-0.25 Hz, scale two corresponded approximately to 0.06-0.125 Hz, and scale three corresponded approximately to 0.06-0.03 Hz. All analyses focused on scale two, which contains the most relevant information for the goals of the present study and has been used in previous studies relating changes in network measures to behavior (e.g., Bassett et al., 2011; Ekman et al., 2012). Although scale one includes frequencies below the Nyquist frequency of 0.25, which could potentially carry information about functional connectivity, the hemodynamic response function acts as a low-pass filter on the underlying neural activity and the signal in high frequencies close to the Nyquist limit can become uninformative (Sun et al., 2004). This is confirmed by experimental estimates showing that both task-related functional connectivity (Richiardi et al., 2011; Sun et al., 2004) and resting-state functional connectivity (Cordes et al., 2001; Richiardi et al., 2011) are found predominately in the low-frequency band of 0.00-0.15 Hz and not in higher frequencies.

To build each functional connectivity matrix, we computed the Pearson correlation coefficient between the wavelet coefficients corresponding to each pair of nodes. We chose a threshold value above which the correlation coefficients were retained and below which the correlation coefficients were set to zero. The threshold was chosen separately for each correlation matrix, in two steps that had the goals of controlling false positive statistical associations and obtaining sparse networks. First, a *t*-test was used to determine whether the correlations deviated significantly from zero and the obtained *p*-values were corrected for multiple comparisons using a false discovery rate of 5% (Benjamini and Hochberg, 1995). The minimal possible threshold was set to the critical value in this omnibus test, which controlled the rate of false positives in each correlation matrix. Second, the value of the threshold was gradually increased in steps of 0.01 and the maximum value yielding a fully connected network was retained. This further reduced the number of edges in the graph, being consistent with the sparsity of anatomical connections in the brain (see Bassett et al., 2006; Meunier et al., 2009). The end result is an adjacency matrix **W** representing a weighted functional network, with elements $W_{ij}$ equal to zero for pairs of nodes with correlations below the threshold and $W_{ij}$ equal to the wavelet correlation for pairs of nodes with correlations above the threshold. In contrast to the common binary network construction, we chose to employ these weighted networks which retain neurophysiologically relevant information about the strength of functional interactions between network nodes (Bassett et al., 2012; Lohse et al., 2013; Rubinov and Sporns, 2011).

*Computation of network theory measures.* We computed six network measures for each functional connectivity matrix using the Brain Connectivity Toolbox for MATLAB (Rubinov and Sporns, 2010): strength, clustering coefficient, characteristic path length, betweenness centrality, intra-modular strength z-score and participation coefficient. These measures were selected from two recent surveys on network diagnostics (Costa et al., 2007; Rubinov and Sporns, 2010) because collectively they parsimoniously capture the local structure of the network surrounding individual network nodes. To facilitate interpretation of results and avoid overly complex regression models, we chose to decrease the number of diagnostics employed by (i) focusing on measures that are commonly used in the neuroscientific literature to facilitate ease of comparison between studies, and (ii) not including multiple measures that provided significantly correlated information.

The *degree* of a node is the number of nodes to which it connects. Specifically, let $a_{ij}$ be a binary variable (i.e., 1 or 0) representing whether or not a connection exists between nodes $i$ and $j$, and let $N$ represent the total number of nodes. Then the degree of node $i$ is computed as:

$$k_i = \sum_{j=1}^{N} a_{ij} \,. \tag{1}$$

Node *strength* is an extension of the definition of degree to weighted networks (Barrat et al., 2004), and is defined as the sum of the weights of a node's connections to other nodes. Let $W_{ij}$ represent the absolute value of the correlation between nodes $i$ and $j$. The strength of node $i$ is:

$$s_i = \sum_{j=1}^{N} W_{ij} \,. \tag{2}$$

The weighted *clustering coefficient* of a node is the fraction of its neighbors that are neighbors of each other. The weighted clustering coefficient of node $i$ can be defined as (Onnela et al., 2005):

$$C_i = \frac{\sum_{j=1}^{N} \sum_{h=1}^{N} \sqrt[3]{W_{ij} W_{ih} W_{jh}}}{k_i \left(k_i - 1\right)} \,. \tag{3}$$

The weighted *characteristic path length* (Newman, 2010) of a node is the average of its weighted shortest path lengths, which are the sum of link lengths separating the node from all other nodes in the network. Let $\gamma_{i\text{-}j}$ represent the set of weighted links along the weighted shortest path from node $i$ to $j$. Then the weighted characteristic path length was computed as:

$$L_i = \frac{\sum_{j=1, j \neq i} \sum_{u,v \in \gamma_{i\text{-}j}} \frac{1}{W_{uv}}}{N - 1} \,. \tag{4}$$

The weighted *betweenness centrality* of node $i$ is the number of shortest paths in the network that pass through node $i$. Let $\rho_{hj}$ represent the number of weighted shortest paths between nodes $h$ and $j$, and let $\rho_{hj}(i)$ represent the number of those paths that pass through node $i$. Then the weighted betweenness centrality of node $i$ is (Freeman, 1978):

$$b_i = \sum_{\substack{h,j \\ h \neq j, h \neq i, j \neq i}} \frac{\rho_{hj}(i)}{\rho_{hj}} . \tag{5}$$

The calculation of additional network diagnostics first required the determination of a partition of network nodes into modules. As in previous research aimed at detecting the community structure of brain networks (e.g., Alexander-Bloch et al., 2012; Bassett et al., 2011; Chen et al., 2008; Meunier et al., 2009), the partition of nodes into modules was found by maximizing a modularity quality function. For a given partition, $g_i$ and $g_j$ represent the modules to which nodes $i$ and $j$ are assigned, respectively. The function $\delta(g_i, g_j)$ is the Kronecker delta, and therefore is equal to one when $g_i = g_j$ and zero otherwise. Then a modularity quality function can be defined as (Newman, 2004):

$$Q = \sum_{i,j} \left[ W_{ij} - \frac{s_i s_j}{\sum_{i,j} W_{ij}} \right] \delta\left(g_i, g_j\right) . \tag{6}$$

We used a Louvain-like locally greedy algorithm (Blondel et al., 2008) to find the partition of nodes into communities that maximized $Q$. Theoretical work (Good et al., 2010) has shown that maximization of $Q$ is complicated by the fact that many different partitions yield near-optimal values of $Q$, together forming a high-modularity plateau in the optimization landscape. To deal with this near-degeneracy, we performed 100 optimizations of the modularity quality function using the Louvain heuristic and we extract a consensus partition from the resulting 100 partitions using the consensus partition method proposed by (Bassett et al., 2013a). Additional results of the analysis of modularity can be found in the supplementary material.

Given a partition of the network into modules, the *intra-modular strength* is the sum of connection weights of a node with nodes from its own module $m$:

$$s_i^m = \sum_{j=1}^{N} W_{ij} \delta\left(g_i, g_j\right) . \tag{7}$$

This value is standardized to obtain an intra-modular strength $z$-score (based on the binary version of this diagnostic first introduced in Guimera and Amaral, 2005):

$$z_i = \frac{s_i^m - \overline{s}^m}{\sigma^m} , \tag{8}$$

where $\overline{s}^m$ and $\sigma^m$ represent the mean and standard deviation (respectively) of the strength distribution for module $m$.

Given a partition of the network into modules, the *participation coefficient* (based on the binary version of this diagnostic first introduced in Guimera and Amaral, 2005) measures the uniformity of the distribution of connections of a node to nodes from all partitions. Values close to one reflect a uniform distribution of connections, and values close to zero reflect a high concentration of connections to only one or a few modules. The measure is defined as:

$$P_i = 1 - \sum_{m=1}^{M} \left( \frac{s_i^m}{s_i} \right)^2 . \tag{9}$$

***Multilevel regression.*** We used multilevel regression (Gelman and Hill, 2007) to explore the relationship between local network dynamics and behavioral changes across blocks of training in the categorization task. These analyses were performed using the package *lme4* (Bates et al., 2014) within the statistical software R v. 3.0.2 (R Development Core Team, 2014). Accuracy and mean correct response time at each training block were used as outcome variables in separate analyses, with node measures as predictor variables.

Because collinearity and multicollinearity of the predictors can have adverse effects on the estimates of regression coefficients (Rawlings et al., 1998; Dormann et al., 2013), we performed a collinearity analysis (see supplementary material) which revealed that characteristic path length and strength were almost perfectly correlated across all nodes (with a median of -0.97 and a range of -0.85 to -0.99). For this reason, we decided to exclude characteristic path length from the regression analysis. The remaining predictors were strength, clustering coefficient, betweenness centrality, intra-modular strength z-score and participation coefficient. The collinearity analysis did not reveal any remaining issues in this set of predictors.

Initial scrutiny of the behavioral data revealed extremely poor performance (near-chance performance and response times larger than 1,000 ms) by a single participant during the first three blocks of training. Because performance jumped to high levels after these three blocks, it is likely that the initial outliers were produced by the introduction of the task in the scanner (e.g., by misunderstanding of instructions). The three outlier data points were removed from all analyses to avoid the influence of a likely artifact on the results.

To determine possible violations of the assumptions underlying linear regression, particularly homoscedasticity and normality of residuals, we performed an analysis of the distribution of residuals after performing an initial regression analysis using the untransformed percent of correct choices and mean response times (see supplementary material). This analysis revealed violations of the assumptions of the linear regression model. These violations were corrected by applying an arcsine-square-root transformation to the accuracy data and a power transformation to the response time data (recommended in Rawlings et al., 1998).

Because our main interest was to explore how local network dynamics are related to behavioral changes across training in the categorization task, block was the main unit of analysis in the model. We used a varying-intercepts model (Gelman and Hill, 2007), which includes a group intercept and individual intercepts for each participant, but only a group regression weight for each predictor. This implements the assumption that different participants showed different baseline levels of performance in the task, but that the relation between network measures and behavior (represented by a single regression coefficient) was the same across participants.

For each node and outcome variable, six models were initially fit to the data: a null model and five explanatory models. It is important to note that model selection does not follow the logic of frequentist null hypothesis significance testing. However, after model selection, we do confirm whether the selected model is better than the null through a likelihood ratio test. This test does follow frequentist logic and is therefore corrected for multiple comparisons.

The null model included only intercepts and no predictors, representing a baseline of performance that was constant across blocks but could vary across participants. This model was used to test whether variations in the predictors across blocks could explain additional variation in behavior. The simplest explanatory model included each of the predictors, but no interaction terms. Although this model had the advantage of being easy to interpret, theoretical considerations suggest that specific combinations of values for two or more predictors could describe the role of a node in the network better than all predictors considered separately. For example, it has been suggested that specific combinations of values for the participation coefficient and intra-modular strength z-score might determine a set of discrete roles for network nodes (Guimera and Amaral, 2005). Because we had no hypotheses about the specific interaction terms that might be important for the prediction of behavior, our approach was to fit a sequence of models of increasing complexity, with each model in the sequence incorporating all interactions one level above those included by the previous model. Thus, one model included all interactions between two predictors, the next model included all interactions between two or three predictors, and so on. The most complex model that we considered included five-predictor interactions.

Each model was fit to the data using maximum likelihood estimation. Model selection was performed using the Akaike information criterion (AIC Akaike, 1974). However, as the AIC is known to be biased for small samples (Burnham and Anderson, 2004), we used a version that corrects for this bias (Hurvich and Tsai, 1989):

$$AIC = -2\log L + 2p\left(\frac{T}{T-p-1}\right), \tag{10}$$

where $L$ is the likelihood of the model at the solution, $p$ is the number of free parameters in the model and $T$ is total number of data points.

Given a set of candidate models indexed by $d = 1, 2, ... D$, it is possible to obtain an estimate of the probability that each of them is the best model in the set, computing $AIC$ weights (Burnham and Anderson, 2004):

$$AICw_d = \frac{\exp\left(-\Delta_d/2\right)}{\sum_{r=1}^{D}\exp\left(-\Delta_r/2\right)}, \tag{11}$$

where

$$\Delta_d = AIC_d - AIC_{\min}, \tag{12}$$

and $AIC_{\min}$ is the lowest $AIC$ value for the set of $D$ models. The best-fitting model for a particular node is the one that produces the smallest AIC and largest AIC$w$.

AIC selects a model with enough complexity to explain as much variability in the behavioral data as possible without overfitting (Burnham and Anderson, 2004). We used this approach to determine: (i) whether including node measures as predictors in the model provided a better description of behavior than the null model, and (ii) whether or not complex models (with second- or higher-order interactions among predictors) would provide a better description of behavior than a simple model assuming independent influences from each predictor.

After model selection, nodes for which the best-fitting model was the null model were removed from the following analyses. When the best-fitting model was not the null, we confirmed that the selected model provided a better fit than the null model by performing a likelihood ratio test. The *p*-values from such tests were corrected for multiple comparisons using a false discovery rate of 5% (Benjamini and Hochberg, 1995). Only nodes in which the best-fitting model provided a significantly better fit than the null model were included in the subsequent analyses. In this way, the outcome of model selection with AIC was supported by an additional criterion based on a traditional statistical test. This procedure resulted in a sample of nodes that were deemed predictive of behavior according to a rather conservative criterion.

***Analysis of regression coefficients.*** Because the best-fitting model in the large majority of predictive nodes did not include interaction terms (see Results section), we were able to focus on such nodes to study the relationship between node measures and behavior as expressed through the regression coefficients from the model. Because better performance is related to higher accuracy but lower response times, we reversed the sign of the regression coefficients from the response time analysis to make them more comparable to those from the accuracy analysis.

In the first part of this analysis, we determined whether the overall distribution of regression coefficients across nodes was biased towards positive or negative values for different node measures. We would expect such biases if a particular node measure was predictive of specific changes in behavior across a majority of nodes. For example, an increase of strength for most connections (i.e., more integrative processing) could be predictive of increments in accuracy. Under the null hypothesis of centered distributions (i.e., with median equal to zero), the number of coefficients larger than zero follows a binomial distribution. We used this fact to statistically determine whether the bias present in each distribution was larger than expected by chance.

Next, we determined whether the distribution of coefficients for a given predictor varied depending on whether the outcome variable was accuracy or response time. As indicated earlier, changes in these two behavioral measures should correlate with different learning mechanisms: accuracy with early category learning and response time with later automaticity learning (Ashby et al., 2007). To compare the two distributions, we performed a permutation test on the absolute difference in median regression coefficient, by randomly re-ordering the outcome label ("accuracy" and "response time") of each regression coefficient 10,000 times and computing the absolute difference in medians after such randomization. This resulted in a distribution of difference values under the null hypothesis that the distribution of regression coefficients was the same for accuracy and response times, which was used to determine whether the observed difference was significantly higher than expected by chance.

Finally, to analyze the distribution of mean AIC*w* and regression coefficients across ROIs, we ranked such values and computed mean rankings for different area types (subcortical, visual, motor and high-level areas). To determine whether these mean rankings were different from what would be expected by chance, we performed permutation tests by randomly re-ordering the area types of all nodes involved in a particular analysis 10,000 times and computing the mean ranking of each area type after such randomization. This resulted in a distribution of mean rankings under the null hypothesis of no effect of area type, which was used to determine whether the observed ranking for each area type was significantly higher than expected by chance. The same distributions were used to determine whether the observed difference between accuracy and response time in mean ranking of each area type was significant.

# Results

## Behavioral results

Figure 1C-D plots behavioral data as a function of training blocks for individual participants and the group mean, demonstrating that performance improved across blocks. Interestingly, mean accuracy (top) reached asymptotic levels around block 7 (second scanner session) while mean response time continued to improve across training. Figure 1E confirms that accuracy and correct response time were weakly correlated across blocks and participants (Pearson's $r = -0.2$, $p < 0.01$), consistent with the possibility that different neurophysiological mechanisms might underlie changes in each behavioral variable.

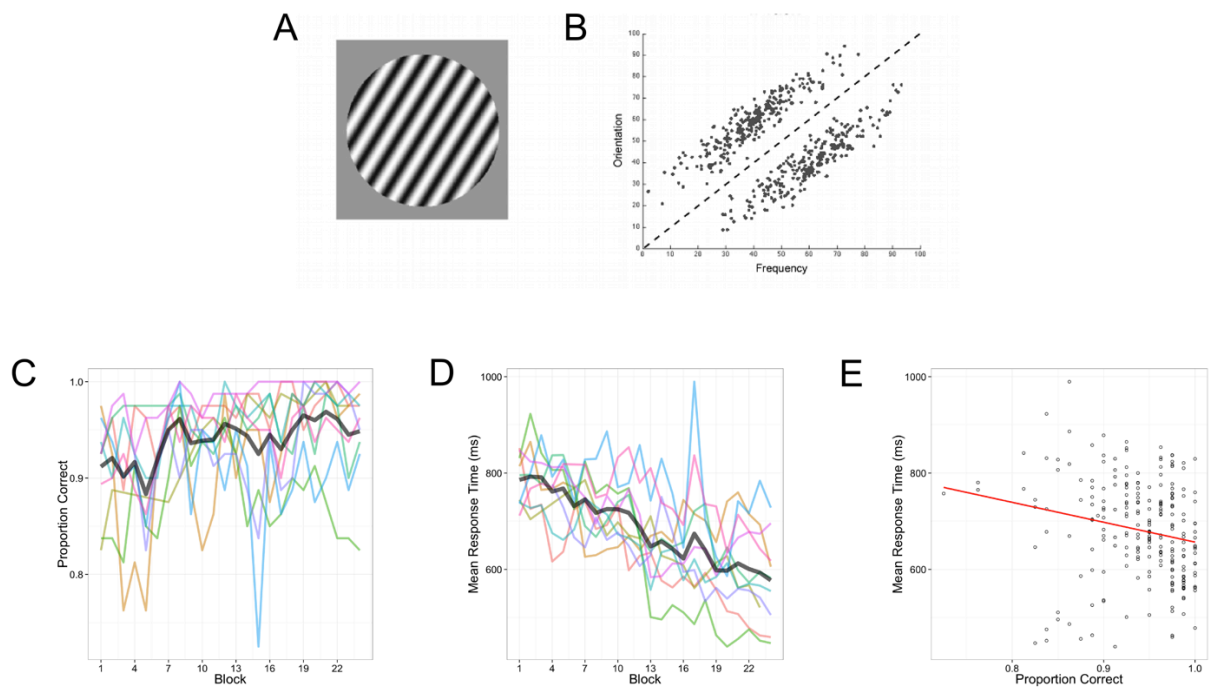

**Figure 1.** Graphical summary of the behavioral task used in this study. (A) Example of a stimulus presented to participants in the categorization task. (B) Representation of the category structure learned by participants. (C) Proportion of correct responses as a function of scanning blocks; lines of different color represent different participants and the grey line represents the group mean. (C) Mean correct response time as a function of scanning block; lines of different color represent different participants and the grey line represents the group mean. (E) Scatterplot of the relationship between proportion of correct responses and correct mean response time across participants and scanning blocks.

## Model selection and distribution of predictive nodes

Out of the 742 nodes in each network, 148 passed our criteria to be considered predictive of changes in accuracy across training, whereas 129 passed our criteria to be considered predictive of changes in response times. The best-fitting models at these nodes had variable complexity, as indexed by the level of interaction terms included in the model. Histograms of the frequency distributions of selected levels of interaction are shown in Figure 2A. For the large majority of nodes, the selected model was one in which no interaction terms were included (level 1), followed by a small proportion of nodes in which the best model included only interactions between pairs of predictors (level 2). Best models with interaction levels larger than 2 were comparatively infrequent. These results suggest that in most cases complex combinations of node measures are not required for the prediction of variations in behavioral measures. Rather, individual node measures can be related to behavioral changes in a straightforward, easily interpretable manner. We therefore focused on those nodes for which the best-fitting model had no interaction terms and we determined the relationship between each node measure and behavior variables through an analysis of regression coefficients (see below).

Figure 2B shows the distribution of AIC$w$ (Eq. 11) for predictive nodes across cortical regions. It can be seen that predictive nodes were distributed evenly across the included cortical areas, with high AIC$w$ in motor, prefrontal and visual areas. The bar plots depict mean AIC$w$ across ROIs, with ROIs ranked from highest to lowest AIC$w$. Nodes in which the best model was the null model were assigned an AIC$w$ of zero. The resulting values summarize how predictive of behavior was each ROI relative to its overall size (i.e., number of nodes in the ROI). Bars are colored according to the type of area they represent, with blue bars representing visual areas, green bars representing motor areas, yellow bars representing areas related to high-level cognition (prefrontal areas and the hippocampus), and red bars representing subcortical areas.

Figure 2B shows that subcortical ROIs were the most highly predictive of behavior in the analysis of accuracy. Areas related to high-level cognition were the least predictive of behavior, whereas motor and visual areas tended to be in the middle of the ranking. A permutation test revealed that the mean ranking of subcortical areas was significantly higher than expected by chance, $\bar{x}$ = 9.83, $p$ <0 .001, the mean ranking of high-level cognition areas was significantly lower than expected by chance, $\bar{x}$ = 26.5, $p$ < 0.01, and the mean ranking of visual ($\bar{x}$ = 17) and motor ($\bar{x}$ = 21.5) areas was not significantly different from chance, $p$ > 0.1.

Figure 2B reveals a similar ranking of areas in the analysis of response times; that is, subcortical areas are ranked relatively high, high-cognition areas are ranked relatively low, and motor and visual areas fall between the two extremes. However, the separation of rankings by area type is less clean that that observed in accuracy. Permutation tests indicated that, for response times, the mean rankings of visual areas was significantly higher than expected by chance, $\bar{x}$ = 10.75, $p$ < 0.05, whereas the mean rankings of subcortical areas ($\bar{x}$ = 16.5), motor areas ($\bar{x}$ = 19.2), and higher-level cognition areas ($\bar{x}$ = 23.3), were not significantly different from what would be expected by chance (all $p$ > 0.05). Furthermore, it was found that the mean ranking of subcortical areas was significantly higher for accuracy than for response times, $p$ < 0.05, but differences in other area types were not significant ($p$ > 0.05).

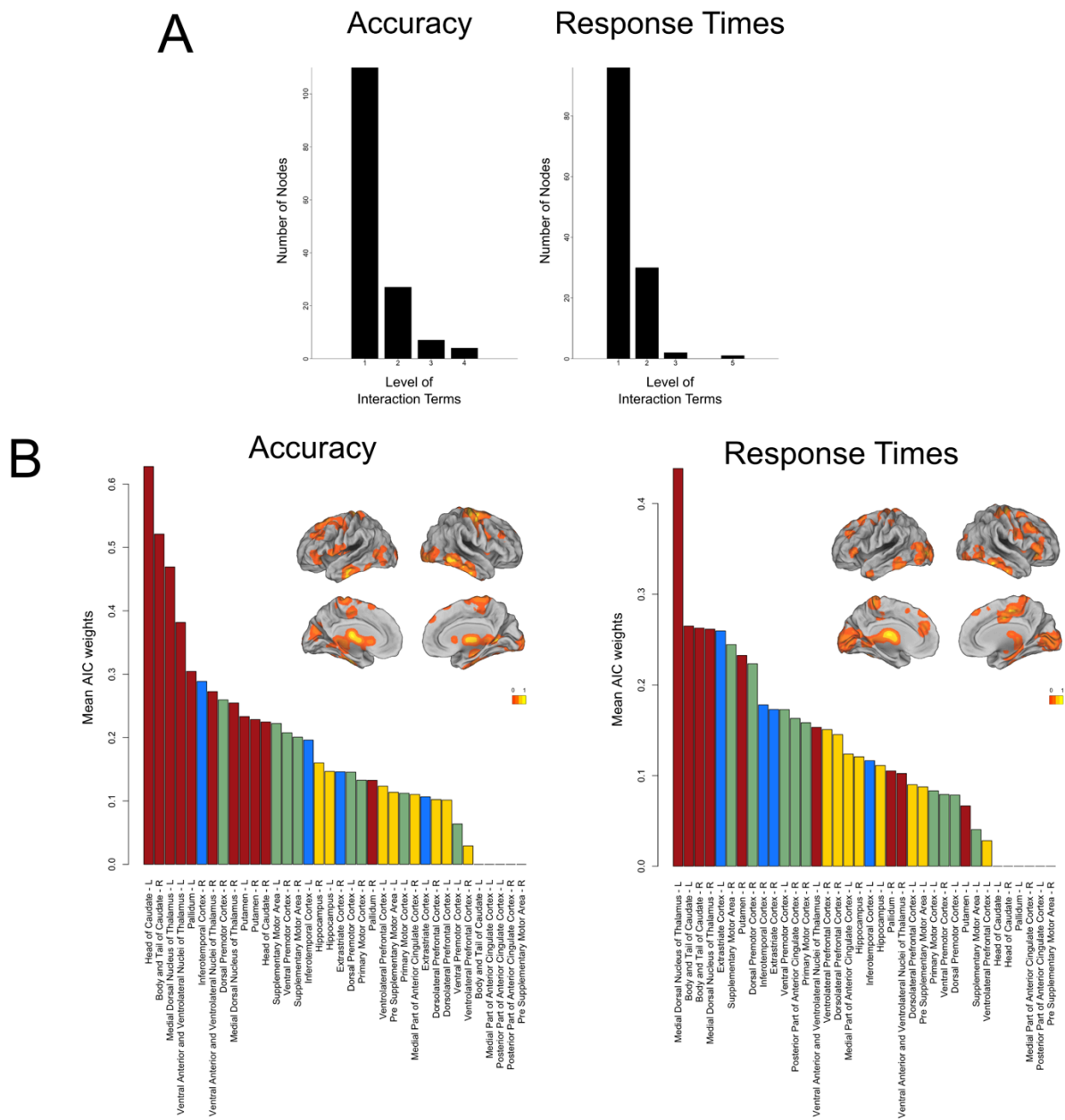

**Figure 2.** Characterization of best-fitting models in predictive nodes in terms of their distribution of complexity and fit to the data. (A) Distribution of the level of interaction terms for the best-fitting model across predictive nodes. (B) Distribution of AIC weights for the best-fitting model of predictive nodes across the cortical surface (brain maps, smoothed using a Gaussian kernel with standard deviation of 3 mm) and regions of interest (bar plots).

## Relation between local network measures and behavior

Next, we were interested in characterizing the direction and magnitude of the relationship between each node measure and behavioral performance, which is possible through exploration of the regression coefficients obtained from the best-fitting multilevel model at each node. The addition of interaction terms in the model makes the interpretation of each predictor difficult. Fortunately, as shown in Figure 2A, for most nodes the best fitting model did not include any interaction terms. We therefore focused on these nodes and ignored the minority of nodes with best-fitting models that included interaction terms.

***Clustering coefficient.*** Figure 3 shows the distributions of standardized regression coefficients for the clustering coefficient, a measure thought in some contexts to relate to the efficiency of local information processing (Bullmore and Sporns, 2009). Panel A shows violin plots with the overall distributions across nodes. The contour of each violin plot represents an estimate of the density of nodes with particular coefficient values, the black bar at the center of each plot represents the interval containing the central 50% of the values in the distribution (bounded by the second and third quartiles), and the white circle inside this bar represents the median of the distribution. Distributions from both the accuracy and response time analysis are clearly unimodal and biased toward positive values (binomial $p < 0.001$ in both cases). This means that, in general, the presence of higher clustering coefficients across the network was related to better performance of the task. The strength of this bias was higher in the analysis of accuracy than in the analysis of response times (permutation test, $p < 0.001$), suggesting that the clustering coefficient was a stronger predictor of accuracy than of response time.

Figure 3B shows the mean value of regression coefficients for each region of interest, which indicates where in the brain a clustering coefficient was a positive or negative predictor of task performance. In the case of accuracy, almost all of the ROIs show a positive mean value (only left extrastriate cortex has a mean slightly below zero). All area types–each represented with a different bar color–tend to include ROIs with high and low rankings. Permutation tests indicated that mean ranking was not significantly different from what would be expected by chance in any area type.

In the case of response times, only two ROIs show negative mean values (right posterior ACC and right SMA). For all other ROIs, higher clustering coefficients were associated with faster response times on average. Subcortical areas (red bars) tended to have high rankings on average (permutation test: $\bar{x} = 8.11$, $p < 0.01$). All other permutation tests were not significant. Furthermore, the mean ranking of subcortical areas was significantly higher for response time than for accuracy, $p < 0.01$, whereas the mean ranking of motor areas was significantly higher for accuracy than for response time, $p < 0.05$. No other differences in mean ranking were significant.

In sum, the relation between the clustering coefficient and task performance was quite straightforward: better performance in terms of both accuracy and speed was accompanied by higher clustering across all area types, with the relation between clustering and speed being particularly strong in subcortical areas.

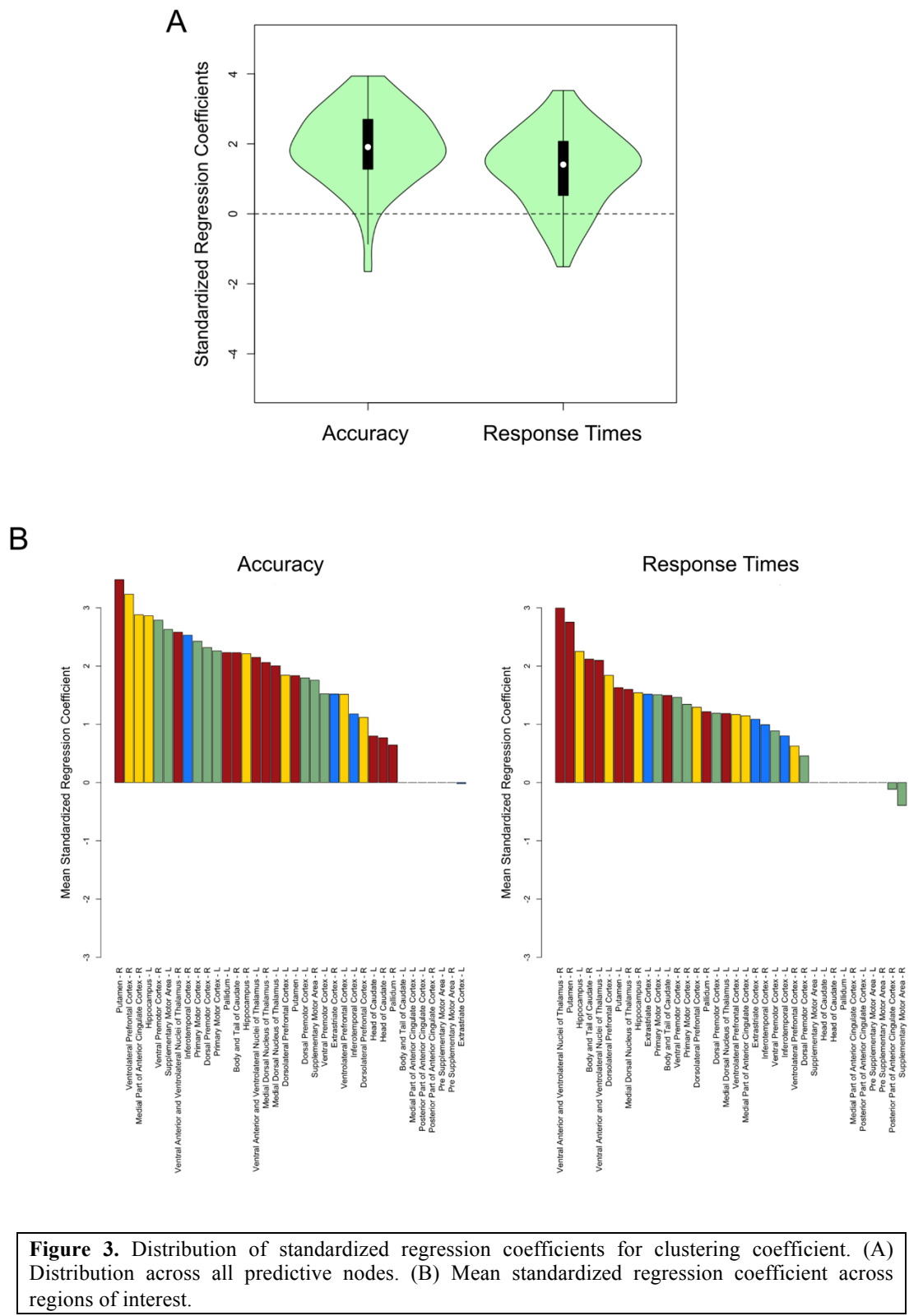

**Figure 3.** Distribution of standardized regression coefficients for clustering coefficient. (A) Distribution across all predictive nodes. (B) Mean standardized regression coefficient across regions of interest.

**Strength.** Figure 4 shows the distributions of standardized regression coefficients for strength, which measures the magnitude of functional connectivity present between nodes in an ROI and the rest of the brain. The violin plots in Figure 4A indicate that the distribution of strength was biased towards negative values when the behavioral outcome measure was accuracy (binomial $p < 0.001$) and towards positive values when the behavioral outcome measure was response time (binomial $p < 0.01$). The difference between the two distributions was statistically significant according to a permutation test, $p < 0.01$.

As shown in Figure 4B, an average negative relationship between accuracy and strength was found only in four ROIs (left extrastriate cortex, right head of the caudate, right SMA and right dorsolateral PFC) and in all cases the mean regression coefficients had a relatively small absolute value. Visual areas (blue bars) showed significantly lower values than expected by chance (permutation test: $\bar{x} = 6.0$, $p < 0.01$). The mean rankings for all other area types were well within the distribution of values expected by chance (permutation test, $p > 0.05$).

In the case of response times, on average an increase in speed was accompanied by an increase in strength in most ROIs, but there are several exceptions. The mean ranking of each area type was not significantly different from what is expected by chance (permutation test, $p > 0.05$). The mean ranking of visual areas was significantly lower in the accuracy analysis than in the response time analysis (permutation test, $p < .05$).

In sum, an increase in accuracy was predicted by a decrease in the connectivity of nodes across the whole network, except for visual areas. On the other hand, an increase in speed was accompanied by an increase in the connectivity of most, but not all nodes. There were eight ROIs that showed on average the reverse relationship between speed and strength (see Figure 4B) and they were distributed across all area types. These findings suggest that accuracy and speed are differentially driven by stronger versus weaker connectivity between nodes: accuracy requires a cost-efficient decrease in strength whereas speed requires an increase in strength.

**Participation coefficient.** A weaker form of dissociation between accuracy and response times was found for all other predictors, with the distribution of regression coefficients being biased for one behavioral variable, but not the other. As seen in Figure 5A, in the case of the participation coefficient, the distribution is biased toward positive values when the outcome is accuracy, $p < 0.001$ and negative values when the outcome is response time, $p > 0.50$. The difference between the two distributions was not significant, $p > 0.05$.

Figure 5B confirms that, on average, an increase in accuracy was accompanied by an increase in participation coefficient across practically all ROIs, with this relation being particularly strong in subcortical areas and SMA. Subcortical areas (red bars) had a mean ranking significantly higher than expected by chance (permutation test: $\bar{x} = 9.82$, $p < 0.01$). Permutation tests for other area types were not significant. In contrast, the direction of the relationship between speed and participation coefficient depended on each specific ROI. Area type did not have an influence in the direction of this relationship, as revealed by non-significant permutation tests. The mean ranking of visual areas was significantly lower for response time than for accuracy (permutation test, $p < 0.05$), but other differences were not significant.

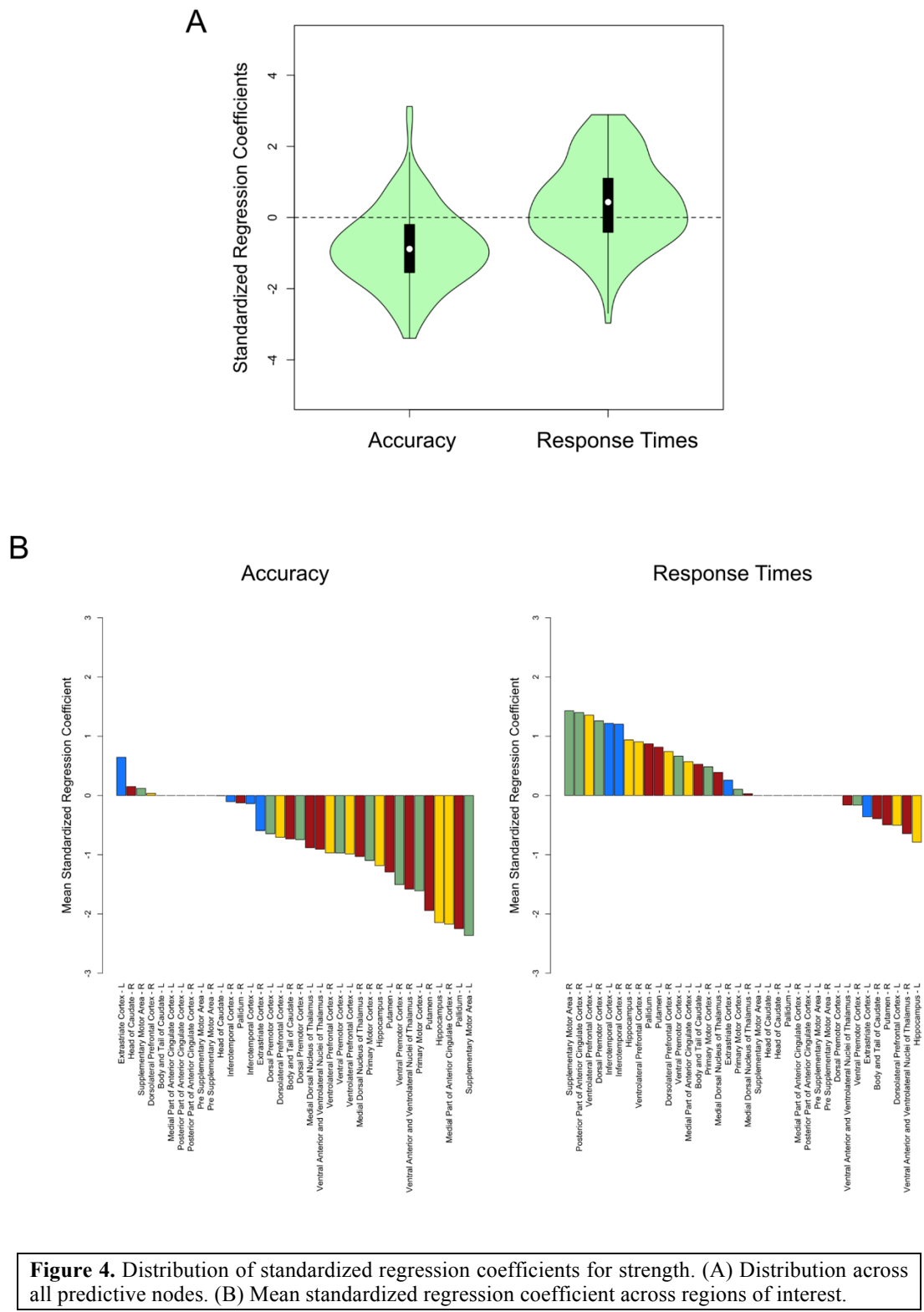

**Figure 4.** Distribution of standardized regression coefficients for strength. (A) Distribution across all predictive nodes. (B) Mean standardized regression coefficient across regions of interest.

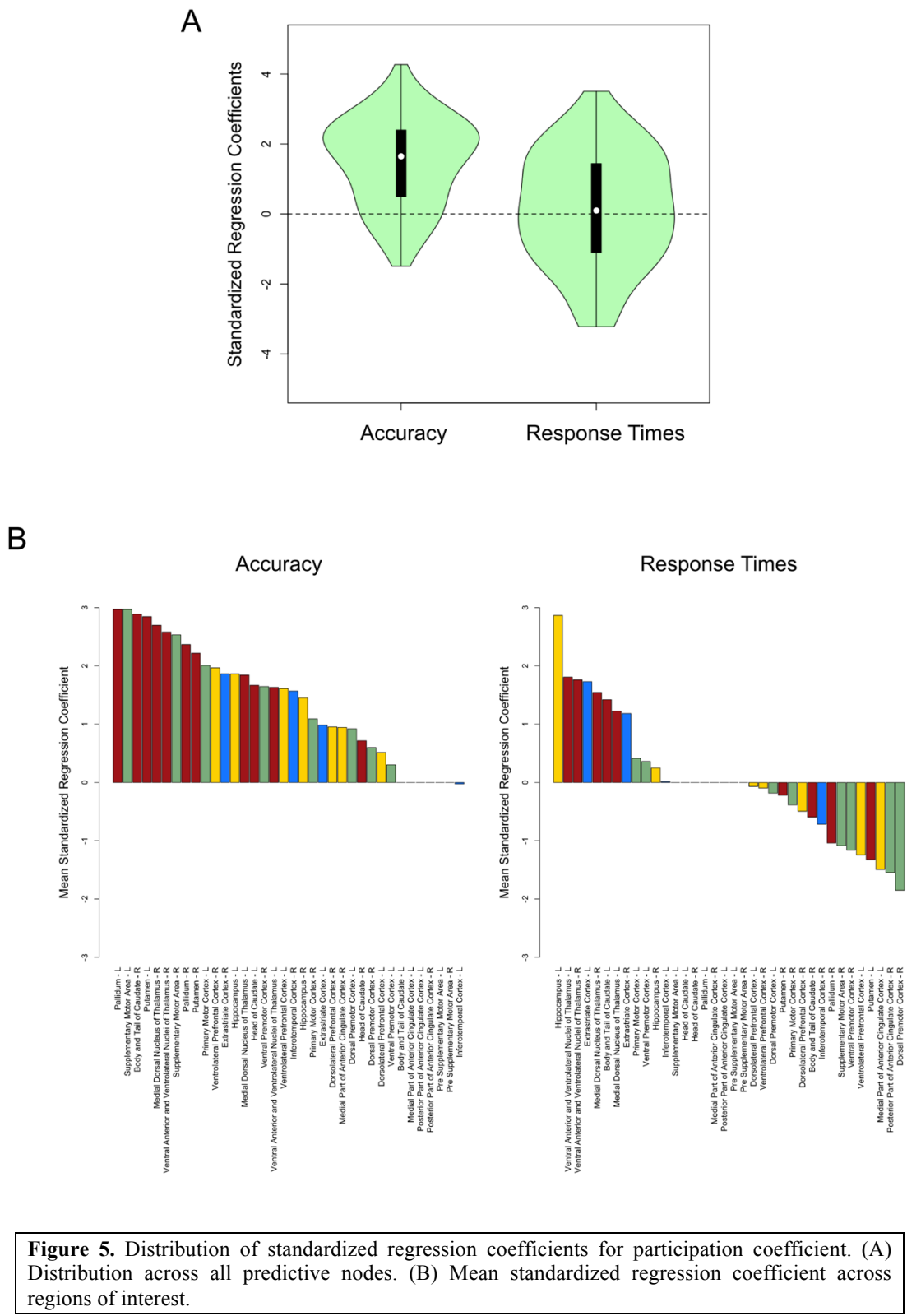

**Figure 5.** Distribution of standardized regression coefficients for participation coefficient. (A) Distribution across all predictive nodes. (B) Mean standardized regression coefficient across regions of interest.

***Intra-modular strength z-score.*** The distribution of regression coefficients for intra-modular strength z-score in Figure 6A shows a pattern of results quite similar to that found for the participation coefficient: the distribution is biased toward positive values when the outcome is accuracy ($p < 0.05$) and negative values when the outcome is response time ($p > 0.10$) and the difference between both distributions was not significant ($p > 0.05$). Note that this qualifies the previously described negative relationship between strength and accuracy: higher accuracy was predicted by lower strength of connections across the network, but by higher relative strength of intra-modular connections, indicating a reorganization of network geometry.

Figure 6B shows that higher accuracy was accompanied by higher intra-modular strength z-score in all subcortical (red bars) and motor (green bars) ROIs, and by lower intra-modular strength z-score in some visual (blue bars) and high-level (yellow bars) ROIs. The mean ranking of subcortical areas was significantly higher than expected by chance (permutation test: $\bar{x} = 10.82$, $p < 0.05$), whereas the mean ranking of visual areas was significantly lower than expected by chance ($\bar{x} = 25.5$, $p < 0.01$). Other tests were not significant.

From the response time analysis, the distribution of mean values across ROIs appears less straightforward. Permutation tests on mean rankings were not significant for any area type (all $p > 0.05$). The mean ranking of visual areas was significantly lower in the analysis of accuracy than in the analysis of response time, $p < 0.01$, but other differences were not significant.

In sum, since the distribution of regression coefficients for both participation coefficient (Figure 5A) and intra-modular strength z-score (Figure 6A) showed a positive bias in the analysis of accuracy, higher accuracy was accompanied by nodes having a connection profile increasingly similar to that of connector hubs (Guimera and Amaral, 2005), which are nodes that allow communication between functionally segregated network communities. This was true for most ROIs, with the exception of visual and some high-level areas (Figure 6B), in which higher accuracy was accompanied by a connection profile increasingly similar to that of connector non-hubs. The relationship between the role of a node in the network's community structure and speed in the task was idiosyncratic to each specific node and was not influenced by area type.

***Betweenness centrality.*** Betweenness centrality showed a pattern opposite to that of the previous two predictors, with the distribution of regression coefficients (Figure 7A) being biased towards negative values in the analysis of response times, $p < 0.001$, but not biased in the analysis of accuracy, $p > 0.50$. The difference between distributions was significant, $p < 0.05$.

The distribution of mean coefficients across ROIs (Figure 7B) reveals that only nine ROIs showed a slightly positive average relation between speed in the task and betweenness centrality. The mean rankings of all area types were not significantly different from what would be expected by chance.

The average relationship between accuracy and betweeness centrality depended on each specific ROI and did not seem to be consistent for any area type. All permutation tests on mean ranking of different area types were not significant ($p > 0.10$). Tests on the difference in mean ranking for each area type between accuracy and response times were also not significant ($p > 0.10$).

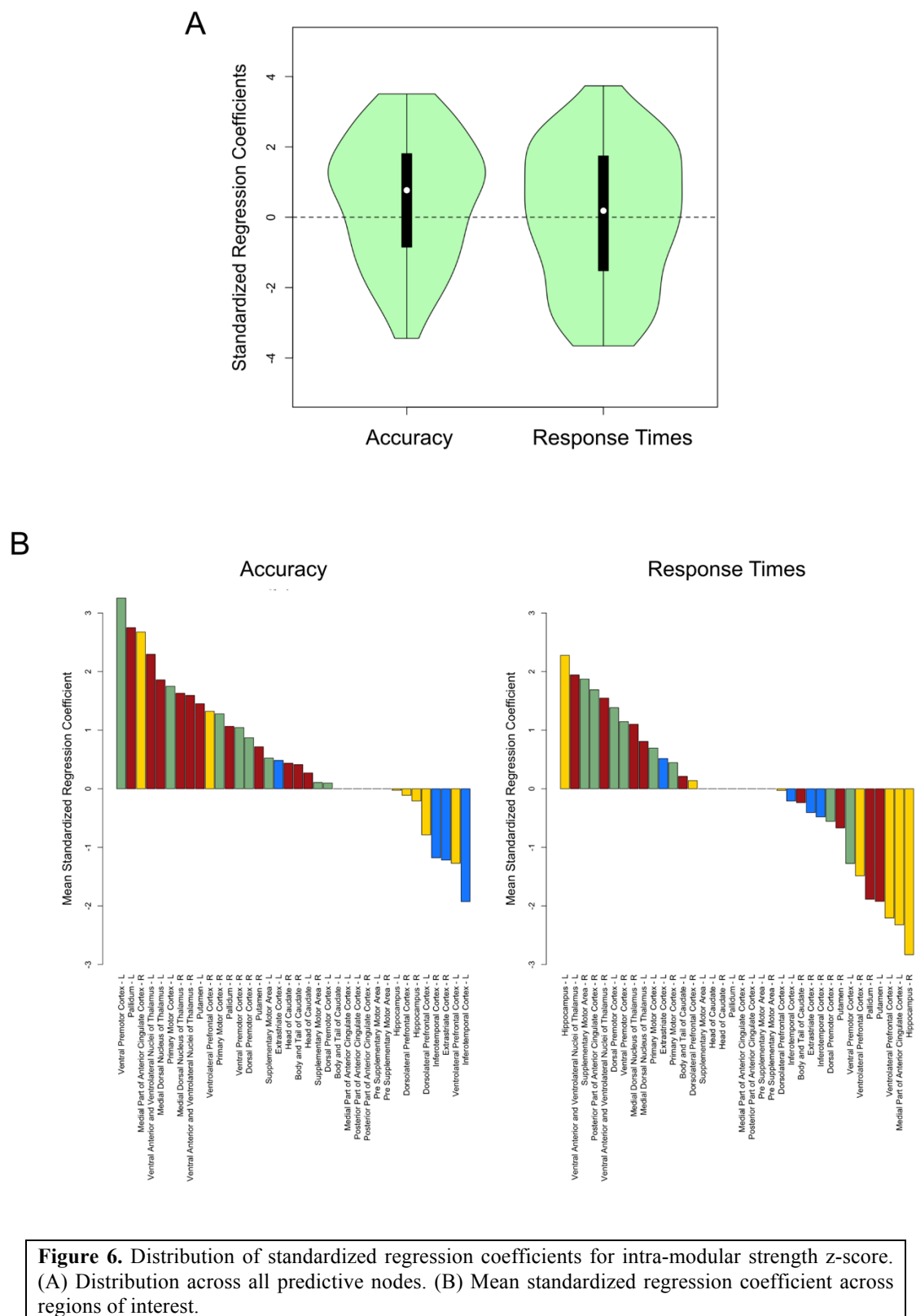

**Figure 6.** Distribution of standardized regression coefficients for intra-modular strength z-score. (A) Distribution across all predictive nodes. (B) Mean standardized regression coefficient across regions of interest.

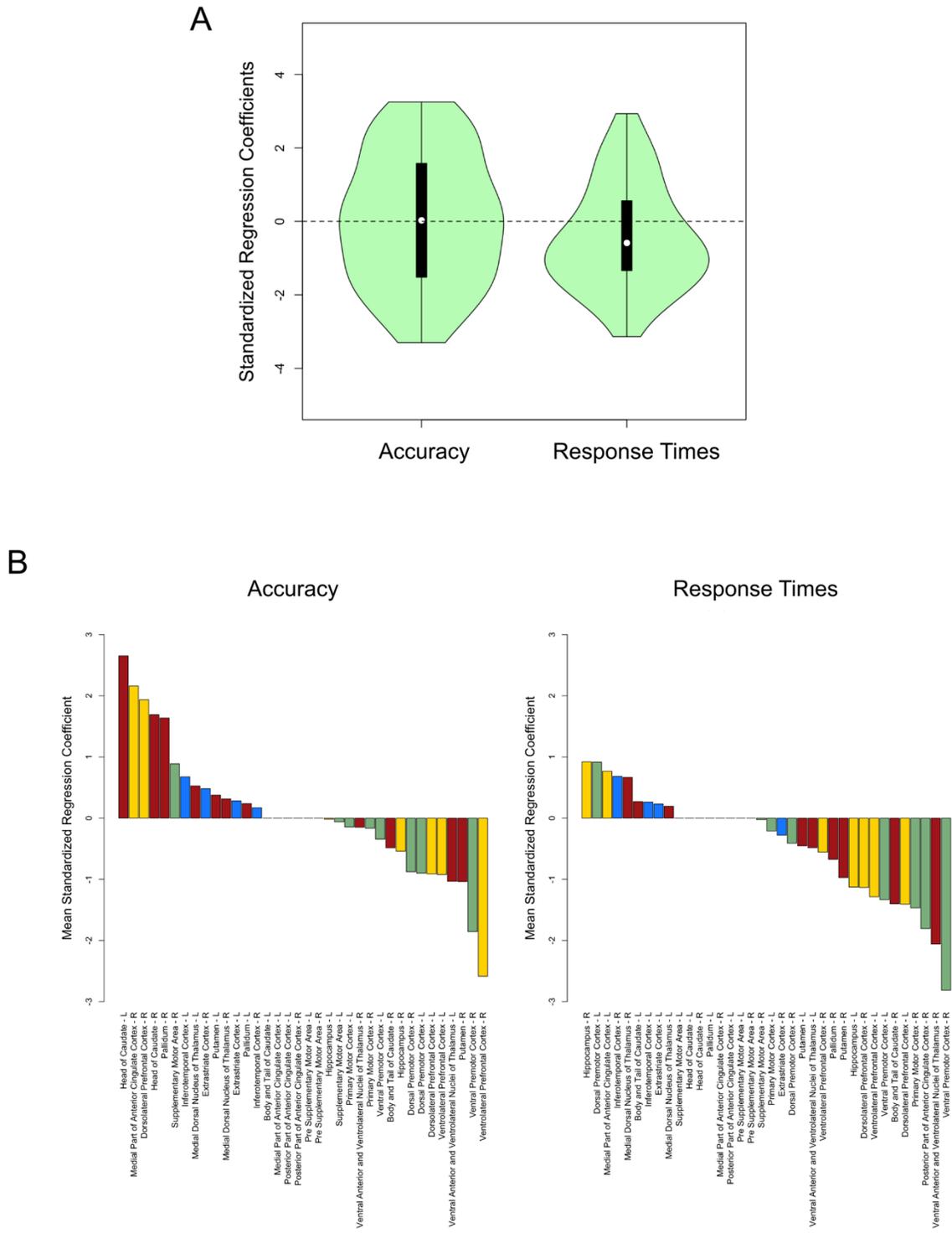

**Figure 7.** Distribution of standardized regression coefficients for betweenness centrality. (A) Distribution across all predictive nodes. (B) Mean standardized regression coefficient across regions of interest.

**Head motion analysis.**

Several recent studies have shown that in-scanner head motion artificially modifies measures of functional connectivity in resting-state fMRI studies and can affect measures derived from network theory (Power et al., 2012, 2014; Satterthwaite et al., 2012; Van Dijk et al., 2012). To account for a possible effect of head motion on our results, we repeated our analyses after removing the influence of head motion from the behavioral variables.

We computed the mean relative displacement (MRD; Satterthwaite et al., 2013) in each functional scan from estimates of head motion provided by MCFLIRT (Jenkinson et al., 2002). We used multilevel regression models with the MRDs as predictors to regress-out the influence of motion from the values of behavioral variables (accuracy and response time). Regressing-out summary statistics of motion in the group-level analysis has been proposed as a way to control for the effects of motion that persist after preprocessing at the individual level (Power et al., 2014; Satterthwaite et al., 2012). We then repeated all our analyses using these motion-corrected outcome variables. We found no indication that a motion artifact could drive the results reported here, which were reproduced in the analysis of motion-corrected data (for more details, see supplementary material).

## Discussion

Our results demonstrate that it is possible to successfully explain variation in performance across training in a categorization task from the geometric network properties of small regions in categorization-related brain areas. As expected, we found that the geometric network properties of nodes in subcortical areas were particularly informative about changes in accuracy, but much less informative about changes in speed of correct responses. This observation is in line with neurocomputational models of automaticity in categorization (see Ashby et al., 2007), which predict that changes in accuracy are due to initial learning supported by basal ganglia, whereas changes in speed of correct responses are largely due to the development of direct connections between sensory and motor areas as skills become automatic.

The best-fitting models for most nodes did not include interaction terms, allowing a straightforward interpretation of the relationship between network predictors and behavior. Increments in accuracy were predicted by a higher clustering coefficient, participation coefficient and intra-modular strength z-score, and by lower strength across the network. The connection strength of nodes within visual areas, both across the whole network and within modules, did not show the same relationship with accuracy as in other areas. On the other hand, speed of correct responses was predicted by higher clustering, higher strength and lower betweenness centrality. This pattern of changes is strikingly similar to that reported in prior studies of motor learning (Heitger et al., 2012), suggesting that these changes in network properties may form a generalizable mechanism of behavioral change across all highly-practiced procedural tasks.

With the exception of the clustering coefficient, we observed a dissociation between accuracy and response times regarding which network properties could consistently explain behavioral changes. For example, higher strength predicted lower accuracy but higher speed of correct responses. This dissociation is in line with the idea that different learning mechanisms are involved in changes of accuracy and speed throughout categorization training. However, the specific relationships found between network measures and behavior are not predicted by neurocomputational theories of categorization and are easier to interpret within the more general framework of network science.

**Effortful processing, automaticity and network geometry**

Recent work has linked effortful information processing during working memory (Kitzbichler et al., 2011) and preparatory attention (Ekman et al., 2012) to higher integration and lower segregation in functional networks. Neurocomputational models (e.g., Ashby et al., 1998) also suggest higher integration during initial category learning in our task, but orchestrated by the basal ganglia and related subcortical structures. The results both support and qualify this hypothesis: learning is predicted by a form of integration that accounts for the network's community structure. Higher accuracy was predicted by higher integration within and across network modules–as measured by the intra-modular strength and participation coefficient, respectively. These changes at the meso-scale were accompanied by lower integration (strength) and higher segregation (clustering) at the global scale of the whole network.

Nodes with a profile increasingly similar to connector hubs predicted higher accuracy. Connector hubs allow communication between functionally segregated network communities and their presence in prefrontal cortex has been previously linked to performance of a motor task (Bassett et al., 2006). It is intuitively plausible that the increased presence of connector hubs enables flexible reconfiguration of brain networks, thereby facilitating learning (Bassett et al., 2011, 2013b). Here, task-related connector hubs were particularly noticeable in subcortical areas, which is consistent with the nature of our task (see Ashby and Spiering, 2004; Ashby and Ennis, 2006).

Taking speed of correct choices as an index of automaticity learning, we found that automaticity is related to increased segregation, as indexed by the clustering coefficient, a relationship that was particularly strong in subcortical areas. This supports our hypothesis that the basal ganglia and related subcortical areas are involved in integrative processing early in category learning, but are released from such function with overtraining. Previous findings of decrements in dynamic centrality with learning of a motor task have also been interpreted as arising from a lower requirement for integration as a skill becomes automated (Mantzaris et al., 2013). Furthermore, the rate of decreased integration between motor and visual modules in a finger sequencing task practiced over 6 weeks predicted individual differences in learning, further highlighting the utility of network segregation for automatic processing (Bassett et al. 2011).

Although automaticity learning was also predicted by an increase in strength in many ROIs, this effect appeared stronger in visual and motor areas than in subcortical areas (Figure 4B). Strength is a measure of the connectivity of pairs of regions, which is expected to increase in cortical areas as automaticity develops. On the other hand, integration involving more than two regions, as measured by betweenness centrality, did decrease with automaticity.

**Focal points of functional network reconfiguration are task-specific.**

Several recent papers have shown that multimodal association areas–including frontal, temporal and parietal cortex–are focal points of functional network reconfiguration during learning and the performance of cognitive tasks (Bassett et al., 2006, 2009, 2011, 2013b; Ekman et al., 2012; Fornito et al., 2012; Cole et al., 2013). Here, the network geometry of frontal and medial temporal nodes played only a minor role in explaining behavioral changes, whereas subcortical areas played a major role. One likely explanation for this departure from previous studies is that they used tasks that engage explicit processes, such as working memory (Bassett et al., 2009), preparatory attention (Ekman et al., 2012), memory recollection (Fornito et al., 2012) and rule use (Cole et al., 2013). On the other hand, we used a categorization task known to engage implicit procedural learning (Ashby et al., 2003; Maddox et al., 2004) mediated by the basal ganglia (Ashby and Spiering, 2004; Ashby and Ennis, 2006).

It is unlikely that a single set of brain regions drives functional network reconfigurations across all tasks. However, some general principles might still be at work. For example, areas with structural connections to a variety of other areas across the brain are more likely to play a role in integrative function (Sporns, 2013, 2014; van den Heuvel and Sporns, 2013). Both multimodal association areas in cortex and the basal ganglia fit this description. The basal ganglia in particular receive projections from most of the cortex and project back to it through the thalamus (Haber and Johnson Gdowski, 2004), which might explain their strong functional connections to key cortical hubs (Zuo et al., 2012) and their role in learning and cognitive function (e.g., Ashby and Ennis, 2006; Ding and Gold, 2013; Doyon et al., 2009; Packard and Knowlton, 2002).

## Methodological considerations

It has been proposed that pre-selecting a small subset of graph measures for analysis is a questionable practice (Ekman et al., 2012). Here we chose a small number of measures because one of our main goals was to interpret the relationship between graph measures and behavior, as revealed by regression coefficients. Including all possible measures makes the interpretation of individual weights impossible due to collinearity issues (Dormann et al., 2013; see supplementary material).

Here we study networks built using functional connectivity measures, which are measures of statistical association between signals in pairs of nodes. Measures of functional connectivity may or may not reflect the strength of information transmission between two areas, as the former can change without changes in the latter (Friston, 2011; Horwitz, 2003). However, the interpretation of our results does not change significantly when we take these issues into consideration. For example, correlated signals suggest that two areas, if not directly connected, at least process similar information or process information in a similar way. Network measures can still be interpreted in terms of integration and segregation of function, and integration may be more costly than segregation if it utilizes redundant information processing.

On the other hand, an area may change functional connectivity without any underlying change in effective connectivity (Friston, 2011), meaning that localization of an effect in a particular ROI suggests, but does not imply, localization of the underlying mechanisms in that ROI. Finally, relations between functional connectivity changes and behavioral changes should not be interpreted as causal, since both are likely the product of latent changes in effective connectivity.

## Conclusions

Using network science and multilevel regression, we showed that changes in the connectivity of small regions can predict behavioral changes during training in a visual categorization task. As expected, we found that changes in the connectivity of subcortical areas were particularly critical for predicting changes in accuracy reflecting initial learning. Such initial learning was predicted by increasingly efficient integrative processing in subcortical areas, with higher functional specialization, more efficient integration across modules, but at a lower cost in terms of redundancy of information processing. Changes in speed of correct responses, reflecting automaticity learning, were predicted by lower clustering (particularly in subcortical areas), higher strength (highest in cortical areas) and higher betweenness centrality. These results support and bring together two recent theoretical developments in computational cognitive neuroscience: neurocomputational theories of category learning and automaticity (Ashby et al., 1998, 2007) and the application of network science to understanding brain connectivity (Bullmore and Sporns, 2009; Sporns, 2014).

# References

Achard S, Salvador R, Whitcher B, Suckling J, Bullmore E (2006) A resilient, low-frequency, small-world human brain functional network with highly connected association cortical hubs. J Neurosci 26:63 –72.

Akaike H (1974) A New Look at the Statistical Model Identification. IEEE Trans Autom Control 19:716–723.

Alexander-Bloch A, Lambiotte R, Roberts B, Giedd J, Gogtay N, Bullmore E (2012) The discovery of population differences in network community structure: New methods and applications to brain functional networks in schizophrenia. NeuroImage 59:3889–3900.

Andersson JLR, Jenkinson M, Smith S (2007) Non-linear registration, aka spatial normalisation. Oxford, UK: FMRIB Analysis Group of the University of Oxford. Available at: http://fmrib.medsci.ox.ac.uk/analysis/techrep/tr07ja2/tr07ja2.pdf [Accessed August 12, 2014].

Ashby FG, Alfonso-Reese LA, Turken AU, Waldron EM (1998) A neuropsychological theory of multiple systems in category learning. Psychol Rev 105:442–481.

Ashby FG, Ell SW, Waldron EM (2003) Procedural learning in perceptual categorization. Mem Cognit 31:1114–1125.

Ashby FG, Ennis JM (2006) The role of the basal ganglia in category learning. In: The Psychology of Learning and Motivation: Advances in Research and Theory (Ross BH, ed), pp 1–36. New York, NY: Elsevier.

Ashby FG, Ennis JM, Spiering BJ (2007) A neurobiological theory of automaticity in perceptual categorization. Psychol Rev 114:632–656.

Ashby FG, Spiering BJ (2004) The Neurobiology of Category Learning. Behav Cogn Neurosci Rev 3:101 –113.

Barrat A, Barthélemy M, Pastor-Satorras R, Vespignani A (2004) The architecture of complex weighted networks. Proc Natl Acad Sci U S A 101:3747–3752.

Bassett DS, Bullmore E, Meyer-Lindenberg A, Apud JA, Weinberger DR, Coppola R (2009) Cognitive fitness of cost-efficient brain functional networks. Proc Natl Acad Sci 106:11747–11752.

Bassett DS, Meyer-Lindenberg A, Achard S, Duke T, Bullmore E (2006) Adaptive reconfiguration of fractal small-world human brain functional networks. Proc Natl Acad Sci 103:19518.

Bassett DS, Nelson BG, Mueller BA, Camchong J, Lim KO (2012) Altered resting state complexity in schizophrenia. NeuroImage 59:2196–2207.

Bassett DS, Porter MA, Wymbs NF, Grafton ST, Carlson JM, Mucha PJ (2013a) Robust detection of dynamic community structure in networks. Chaos Interdiscip J Nonlinear Sci 23:013142.

Bassett DS, Wymbs NF, Porter MA, Mucha PJ, Carlson JM, Grafton ST (2011) Dynamic reconfiguration of human brain networks during learning. Proc Natl Acad Sci 108:7641-7646.

Bassett DS, Wymbs NF, Rombach MP, Porter MA, Mucha PJ, Grafton ST (2013b) Task-based core-periphery organization of human brain dynamics. PLoS Comput Biol 9:e1003171.

Bates DM, Maechler M, Bolker B, Walker (2014) lme4: Linear mixed-effects models using Eigen and S4. Available at: http://cran.r-project.org/web/packages/lme4/index.html.

Benjamini Y, Hochberg Y (1995) Controlling the false discovery rate: A practical and powerful approach to multiple testing. J R Stat Soc Ser B Methodol 57:289–300.

Bell BA, Morgan GB, Schoeneberger JA, Kromrey JD, Ferron JM (2014) How low can you go? An investigation of the influence of sample size and model complexity on point and interval estimates in two-level linear models. Methodol Eur J Res Methods Behav Soc Sci 10:1–11.

Blondel VD, Guillaume JL, Lambiotte R, Lefebvre E (2008) Fast unfolding of communities in large networks. arXiv:08030476 Available at: http://arxiv.org/abs/0803.0476 [Accessed January 6, 2012].

Brainard DH (1997) The psychophysics toolbox. Spat Vis 10:433–436.

Bullmore E, Sporns O (2009) Complex brain networks: graph theoretical analysis of structural and functional systems. Nat Rev Neurosci 10:186–198.

Burnham KP, Anderson DR (2004) Multimodel inference understanding AIC and BIC in model selection. Sociol Methods Res 33:261–304.

Chen ZJ, He Y, Rosa-Neto P, Germann J, Evans AC (2008) Revealing modular architecture of human brain structural networks by using cortical thickness from MRI. Cereb Cortex 18:2374–2381.

Cole MW, Bassett DS, Power JD, Braver TS, Petersen SE, Cole MW (2014) Intrinsic and task-evoked network architectures of the human brain. Neuron 83:238–251.

Cole MW, Reynolds JR, Power JD, Repovs G, Anticevic A, Braver TS (2013) Multi-task connectivity reveals flexible hubs for adaptive task control. Nat Neurosci 16:1348–1355.

Cordes D, Haughton VM, Arfanakis K, Carew JD, Turski PA, Moritz CH, Quigley MA, Meyerand ME (2001) Frequencies contributing to functional connectivity in the cerebral cortex in "resting-state" data. Am J Neuroradiol 22:1326–1333.

Cornish C (2006) WMTSA wavelet toolkit for MATLAB. Available at: http://www.atmos.washington.edu/wmtsa/.

Costa LF, Rodrigues FA, Travieso G, Villas Boas PR (2007) Characterization of complex networks: A survey of measurements. Adv Phys 56:167–242.

DeGutis J, D'Esposito M (2009) Network changes in the transition from initial learning to well-practiced visual categorization. Front Hum Neurosci 3:44.

Dormann CF, Elith J, Bacher S, Buchmann C, Carl G, Carré G, Marquéz JRG, Gruber B, Lafourcade B, Leitão PJ (2013) Collinearity: A review of methods to deal with it and a simulation study evaluating their performance. Ecography 36:027–046.

Ekman M, Derrfuss J, Tittgemeyer M, Fiebach CJ (2012) Predicting errors from reconfiguration patterns in human brain networks. Proc Natl Acad Sci 109:16714–16719.

Fornito A, Harrison BJ, Zalesky A, Simons JS (2012) Competitive and cooperative dynamics of large-scale brain functional networks supporting recollection. Proc Natl Acad Sci 109:12788–12793.

Freeman LC (1978) Centrality in social networks conceptual clarification. Soc Netw 1:215–239.

Friston KJ (2011) Functional and effective connectivity: A review. Brain Connect 1:13–36.

Gelman A, Hill J (2007) Data Analysis Using Regression And Multilevel/Hierarchical Models. New York, NY: Cambridge University Press.

Good BH, de Montjoye YA, Clauset A (2010) Performance of modularity maximization in practical contexts. Phys Rev E 81:046106.

Guimera R, Amaral LA (2005) Functional cartography of complex metabolic networks. Nature 433:895–900.

Haber SN, Johnson Gdowski M (2004) The basal ganglia. In: The human nervous system, 2nd ed. (Paxinos G, Mai JK, eds), pp 676–719. San Diego, CA: Academic Press.

Heitger MH, Ronsse R, Dhollander T, Dupont P, Caeyenberghs K, Swinnen SP (2012) Motor learning-induced changes in functional brain connectivity as revealed by means of graph-theoretical network analysis. NeuroImage 61:633–650.

Helie S, Waldschmidt JG, Ashby FG (2010) Automaticity in rule-based and information-integration categorization. Atten Percept Psychophys 72:1013–1031.

Horwitz B (2003) The elusive concept of brain connectivity. NeuroImage 19:466–470.

Hurvich CM, Tsai CL (1989) Regression and time series model selection in small samples. Biometrika 76:297–307.

Jenkinson M, Bannister P, Brady M, Smith S (2002) Improved optimization for the robust and accurate linear registration and motion correction of brain images. NeuroImage 17:825–841.

Jenkinson M, Smith S (2001) A global optimisation method for robust affine registration of brain images. Med Image Anal 5:143–156.

Kitzbichler MG, Henson RNA, Smith ML, Nathan PJ, Bullmore E (2011) Cognitive effort drives workspace configuration of human brain functional networks. J Neurosci 31:8259–8270.

Lohse C, Bassett DS, Lim KO, Carlson JM (2013) Resolving structure in human brain organization: identifying mesoscale organization in weighted network representations. ArXiv13126070 Q-Bio Available at: http://arxiv.org/abs/1312.6070 [Accessed August 11, 2014].

Maas CJM, Hox JJ (2005) Sufficient sample sizes for multilevel modeling. Methodol Eur J Res Methods Behav Soc Sci 1:86.

Maddox WT, Bohil CJ, Ing AD (2004) Evidence for a procedural-learning-based system in perceptual category learning. Psychon Bull Rev 11:945–952.

Mantzaris AV, Bassett DS, Wymbs NF, Estrada E, Porter MA, Mucha PJ, Grafton ST, Higham DJ (2013) Dynamic network centrality summarizes learning in the human brain. J Complex Netw 1:83–92.

Martin JH (2003) Neuroanatomy: text and atlas. McGraw-Hill Professional.

Meunier D, Achard S, Morcom A, Bullmore E (2009) Age-related changes in modular organization of human brain functional networks. NeuroImage 44:715–723.

Moussa MN, Vechlekar CD, Burdette JH, Steen MR, Hugenschmidt CE, Laurienti PJ (2011) Changes in cognitive state alter human functional brain networks. Front Hum Neurosci 5:83.

Newman MEJ (2004) Analysis of weighted networks. Phys Rev E 70:056131.

Newman MEJ (2010) Networks: An Introduction. Oxford, UK: Oxford University Press.

Nolte J (2008) The human brain: An introduction to its functional anatomy, 6th ed. Philadelphia: Mosby.

Onnela JP, Saramäki J, Kertész J, Kaski K (2005) Intensity and coherence of motifs in weighted complex networks. Phys Rev E 71:065103.

Percival DB, Walden AT (2000) Wavelet methods for time series analysis. Cambridge, UK: Cambridge University Press.

Petrides M, Pandya DN (2004) The frontal cortex. In: The human nervous system, 2nd ed. (Paxinos G, Mai JK, eds), pp 950–972. San Diego, CA: Academic Press.

Picard N, Strick PL (2001) Imaging the premotor areas. Curr Opin Neurobiol 11:663–672.

Power JD, Barnes KA, Snyder AZ, Schlaggar BL, Petersen SE (2012) Spurious but systematic correlations in functional connectivity MRI networks arise from subject motion. NeuroImage 59:2142–2154.

Power JD, Cohen AL, Nelson SM, Wig GS, Barnes KA, Church JA, Vogel AC, Laumann TO, Miezin FM, Schlaggar BL, Petersen SE (2011) Functional network organization of the human brain. Neuron 72:665–678.

Power JD, Mitra A, Laumann TO, Snyder AZ, Schlaggar BL, Petersen SE (2014) Methods to detect, characterize, and remove motion artifact in resting state fMRI. NeuroImage 84:320–341.

R Development Core Team (2014) R: A language and environment for statistical computing. Vienna, Austria: R Foundation for Statistical Computing. Available at: http://www.R-project.org.

Rawlings JO, Pantula SG, Dickey DA (1998) Applied Regression Analysis: A Research Tool. New York, NY: Springer.

Richiardi J, Eryilmaz H, Schwartz S, Vuilleumier P, Van De Ville D (2011) Decoding brain states from fMRI connectivity graphs. NeuroImage 56:616–626.

Rubinov M, Sporns O (2010) Complex network measures of brain connectivity: Uses and interpretations. NeuroImage 52:1059–1069.

Rubinov M, Sporns O (2011) Weight-conserving characterization of complex functional brain networks. NeuroImage 56:2068–2079.

Rzucidlo JK, Roseman PL, Laurienti PJ, Dagenbach D (2013) Stability of whole brain and regional network topology within and between resting and cognitive states. PLoS ONE 8:e70275.

Satterthwaite TD, Wolf DH, Loughead J, Ruparel K, Elliott MA, Hakonarson H, Gur RC, Gur RE (2012) Impact of in-scanner head motion on multiple measures of functional connectivity: Relevance for studies of neurodevelopment in youth. NeuroImage 60:623–632.

Satterthwaite TD, Elliott MA, Gerraty RT, Ruparel K, Loughead J, Calkins ME, Eickhoff SB, Hakonarson H, Gur RC, Gur RE, Wolf DH (2013) An improved framework for confound regression and filtering for control of motion artifact in the preprocessing of resting-state functional connectivity data. NeuroImage 64:240–256.

Snijders TAB, Bosker RJ (2012) Multilevel Analysis: An Introduction to Basic and Advanced Multilevel Modeling, 2nd ed. Thousand Oaks, CA: SAGE.

Soto FA, Waldschmidt JG, Helie S, Ashby FG (2013) Brain activity across the development of automatic categorization: A comparison of categorization tasks using multi-voxel pattern analysis. NeuroImage 71:284–897.

Sporns O (2013) Network attributes for segregation and integration in the human brain. Curr Opin Neurobiol 23:162–171.

Sporns O (2014) Contributions and challenges for network models in cognitive neuroscience. Nat Neurosci 17:652–660.

Sun FT, Miller LM, D'Esposito M (2004) Measuring interregional functional connectivity using coherence and partial coherence analyses of fMRI data. NeuroImage 21:647–658.

Van den Heuvel MP, Sporns O (2013) Network hubs in the human brain. Trends Cogn Sci 17:683–696.

Van Dijk KRA, Sabuncu MR, Buckner RL (2012) The influence of head motion on intrinsic functional connectivity MRI. NeuroImage 59:431–438.

Varoquaux G, Craddock RC (2013) Learning and comparing functional connectomes across subjects. NeuroImage 80:405–415.

Vogt BA, Hof PR, Vogt LJ (2004) Cingulate gyrus. In: The human nervous system, Second. (Paxinos G, Mai JK, eds), pp 915–949. San Diego: Academic Press.

Waldschmidt JG, Ashby FG (2011) Cortical and striatal contributions to automaticity in information-integration categorization. NeuroImage 56:1791–1802.

Zuo XN, Ehmke R, Mennes M, Imperati D, Castellanos FX, Sporns O, Milham MP (2012) Network centrality in the human functional connectome. Cereb Cortex 22:1862–1875.

# Supplementary Material

## 1. Small sample sizes in multilevel regression

The sample sizes used in this study included 10 participants and 24 observations per participant; each model was fit to a total of 237 data points (three points were excluded from the analysis of a single participant). Although the sample size at the participant level was rather small, it is not below the minimum recommended for multilevel regression (Snijders and Bosker, 2012). Furthermore, simulation work using complex multilevel models has shown that confidence interval coverage, type I (false positive) error and statistical bias are all well controlled with sample sizes in the range 5-20, although statistical power is lost (Bell et al., 2014). Other simulations using simpler models have found that 10 participants and 5 observations per participant produce unbiased estimates of the regression coefficients, but biased standard errors that can cause problems with inferential statistics (Maas and Hox, 2005). Fortunately, our analyses do not focus on inferential statistics for single estimates, but on whole distributions and means of estimates for specific ROIs. Thus, low power and biased standard errors at the level of single estimates do not affect our main conclusions.

## 2. Collinearity analysis

Collinearity refers to the situation in which two or more predictors in a regression model are linearly related. The presence of collinearity is problematic when one of the goals of the regression analysis is to understand the relationship between predictors and the outcome variable. As two or more predictors can work as substitutes of each other, estimation of the regression weights is complicated because they become non-identifiable (Dormann et al., 2013; Rawlings et al., 1998). Given that one of the main goals of our study was to evaluate the relation of node measures and behavioral performance through the analysis of regression coefficients, we conducted a collinearity analysis including the six node measures chosen to be included as predictors in our analysis: strength, clustering coefficient, characteristic path length, betweenness centrality, intra-modular strength z-score, and participation coefficient.

For each node in the network, we computed the Pearson correlation coefficient between each pair of predictors. Because such pairwise correlations cannot detect multicollinearity, we also computed the condition number (CN) from the correlation matrix, which is an accepted overall summary of multicollinearity (Dormann et al., 2013). Let $\lambda_i$ represent the $i$th eigenvalue of the correlation matrix; then the CN is computed as:

$$CN = \max_j \left( \sqrt{\frac{\max_i(\lambda_i)}{\lambda_j}} \right). \tag{S1}$$

A commonly used rule of thumb is that CN > 10 indicates that the estimates of regression coefficients might be affected by collinearity, CN > 30 indicates moderate dependencies and CN > 100 indicates serious collinearity (Rawlings et al., 1998).

The left panel of Figure S1 shows the median correlations between pairs of predictors. Contour lines represent bivariate normal densities matching the observed correlation (Murdoch and Chow, 1996). To aid visualization, negative correlations are plotted in red and positive correlations in blue, with darker shades of color representing higher magnitudes. The most important result is an almost perfect negative correlation between characteristic path length and strength, with a median of *r*=-0.97. The correlation at each node ranged from r=-0.85 to r=-0.99. Such high correlations suggest that these two predictors could be substituted for each other in the regression analysis. We therefore chose to exclude the characteristic path length for two reasons: (i) the correlations between characteristic path length and all other variables had a higher magnitude than those of strength and other variables, and (ii) in functional networks, strength has a more straightforward interpretation than characteristic path length (Rubinov and Sporns, 2010).

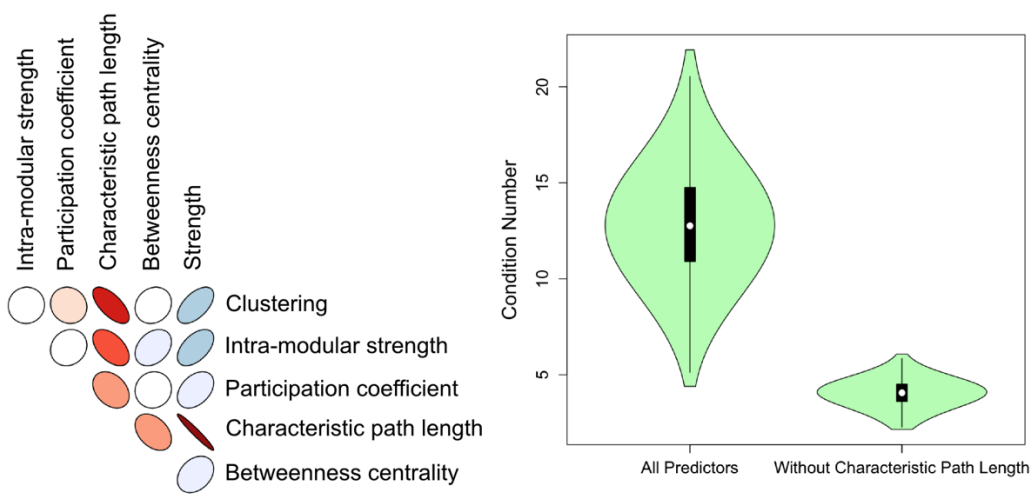

**Figure S1.** Results of the analysis of collinearity of predictors. The left panel represents median correlations between each pair of predictors. Ellipses represent contours of a two-dimensional normal distribution with the same correlation as the two predictors. Negative correlations are represented in red and positive correlations in blue. The right panel summarizes the distribution of condition numbers across nodes before and after elimination of characteristic path length.

The right panel of Figure S1 shows the effect of eliminating characteristic path length on multicollinearity as measured by CN. The figure displays two violin plots, whose contours represent an estimate of CN density across nodes. The black bars at the center of each plot represent the interval containing the central 50% of the values in the distribution (bounded by the second and third quartiles), and the white circle inside this bar represents the median of the distribution. Visual inspection of the CN distribution for all predictors reveals that most nodes (>75%) have CNs above the commonly used threshold of 10, which is the point where multicollinearity begins to be problematic (Rawlings et al., 1998). The CN distribution for all predictors except characteristic path length reveals that elimination of the latter brings all CN values below 10.

It can be seen from Figure S1 that strength also had moderate median correlations with clustering (r=0.57) and intra-modular strength z-score (r=0.52). However, only a small proportion of nodes showed correlations larger than the recommended rule of thumb of r=0.70 (Dormann et al., 2013): in the context of the clustering coefficient, 16% of nodes showed a correlation larger than this rule of thumb while in the context of the intra-module strength z-score, 6% of nodes showed a correlation larger than this rule of thumb. Furthermore, in none of these nodes was the CN larger than 10. From these results, our conclusion is that any collinearity problems introduced by keeping strength as a predictor would be minor, whereas excluding this theoretically important measure might be a serious mistake.

# 3. Residual distribution analysis and transformation of behavioral variables

To determine possible violations of the assumptions underlying linear regression, particularly homoscedasticity and normality of residuals, we performed an analysis of the distribution of residuals after performing the regression analysis using the untransformed percent of correct choices and mean response times.

Using the best-fitting model (see Materials and Methods in main manuscript) from each node, we tested the normality of the residuals through a Shapiro-Wilk test (Shapiro and Wilk, 1965). We tested for possible violations of homoscedasticity by computing the Pearson correlation coefficient between the absolute value of the residuals and the values predicted by the fitted model for each data point. We tested the significance of this correlation against the null hypothesis of no correlation. This correlation test can only detect violations of homoscedasticity in which predicted values and absolute residuals show linear relations; it does not account for non-linear relations (e.g., larger or smaller residuals in the middle of the range of predicted values). Observation of residual plots revealed that such linear relations could be a cause of concern, whereas more complex relations were not apparent. The *p*-values from both Shapiro-Wilks and correlation tests were corrected for multiple comparisons using a false discovery rate of 5% (Benjamini and Hochberg, 1995).

The analysis of residuals when proportion of correct responses was the outcome variable revealed both violations of the assumption of normality and homoscedasticity in 741 of the 742 nodes. The residual-predicted correlations ranged from r=-0.41 to r=-0.11, with a median of r=-0.29. For this reason, we applied an arcsine-square-root transformation to the proportion data, which is usually recommended for making binomially-distributed data closer to the normal and stabilizing its variance (Rawlings et al., 1998). Analysis of residuals after such transformation showed no nodes with significant violations of either the normality or homoscedasticity assumptions. The residual-predicted correlations ranged from r=-0.09 to r=0.10, with a median of r=-0.02. That is, the arcsine-square-root transformation was successful in normalizing the data and stabilizing variances.

The analysis of residuals when mean correct response time was the outcome variable revealed no violations of the assumption of normality, but there were violations of homoscedasticity in 730 of the 742 nodes. The residual-predicted correlations ranged from r=-0.29 to r=0.03, with a median of r=-0.26, meaning that the variance of the residuals increased as the predicted response time decreased. As recommended by Rawlings et al. (1998), we re-ran the analysis using a power transformation of the response time data. It was found that a power transformation with exponent 2 could stabilize the variances across all nodes (residual-predicted correlations ranged from r=-0.16 to r=0.14, with median r=-0.12; no correlations were significant), but introduced violations in the assumption of normality in 697 nodes, according to the Shapiro-Wilk test of normality. We tried several other exponents in the range 1.5-3.0, but values smaller than 2.0 could not stabilize the variance, whereas values higher than 2.0 worsened the violations of normality.

Thus, a decision had to be made between stabilizing variances versus introducing violations of normality of residuals. Variance stabilization was given precedence and power-transformed values were used in the main analysis, for two reasons. First, visual inspection of plots of quantiles from a normal distribution against the quantiles in the distribution of residuals (Q-Q plots) after transformation revealed that violations of normality were small (i.e., points in the Q-Q plot did not deviate importantly from a line). It is likely that such small violations were significant because the large sample size (237) gave high power to the Shapiro-Wilk test.

Second, violations of normality are considered the least important in regression (Gelman and Hill, 2007; Rawlings et al., 1998). The consequence of non-normal residuals is that estimates of standard errors are biased and any inferential statistics performed using such errors are not trustworthy (Maas and Hox, 2004). However, the parameter estimates obtained through maximum likelihood estimation are still consistent and asymptotically unbiased. Importantly, in our analysis of regression coefficients we do not focus on individual statistical tests of significance of coefficients; instead, we focus on whole distributions of coefficients across nodes or on the average coefficient value across regions of interest (see main manuscript).

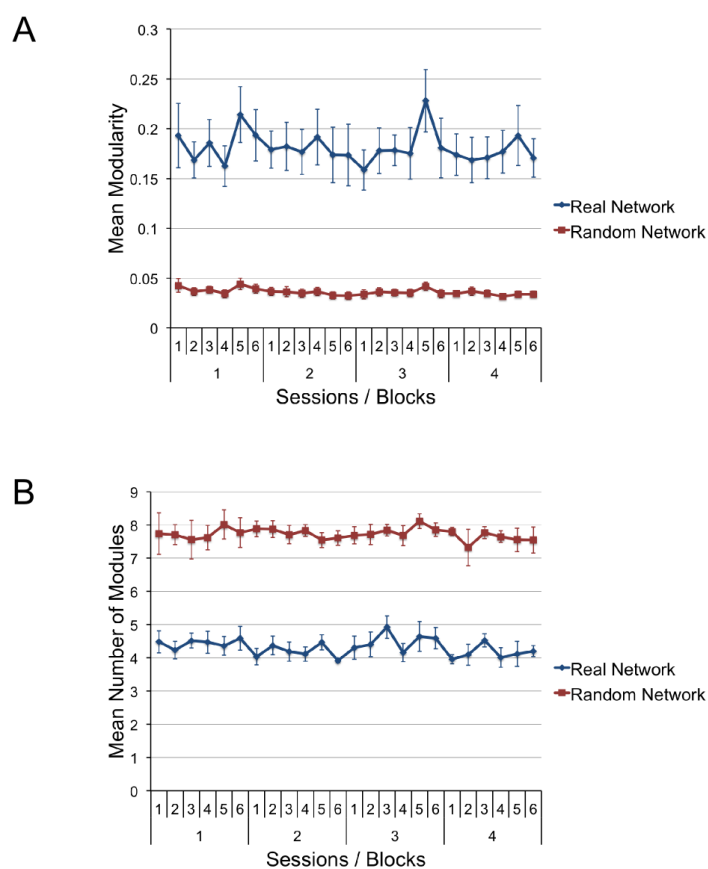

**Figure S2.** Results of the analysis of modularity. (A) Mean modularity index Q of real and random networks across training blocks. (B) Mean number of modules of real and random networks across training blocks. Random networks are constructed to have the same degree distribution as the real networks.

## 4. Modularity analysis

To test whether the modularity quality index Q and number of modules in the networks under study were higher than expected by chance, we built randomized networks with the same degree distribution as the original functional networks (as in Bassett et al., 2011). Each original network was re-wired using the algorithm proposed by (Maslov and Sneppen, 2002). At each step of this randomization procedure, two edges are randomly chosen, one connecting node A to node B and another connecting node C to node D. The edges are re-arranged to connect node A to node C and node B to node D, which preserves the degree of each node, but changes the weight of its connections. The process was repeated multiple times to ensure that the expected number of times an edge was rewired was 20.

For each network, 100 instantiations of this null model were built and the community structure of each random network was obtained using the same procedure used for real networks (see Methods in main report).

Figure S2 summarizes the results of our analysis of modularity in the functional networks and its change across training. Panel A shows that the mean value of modularity for real networks was between Q=0.15 and Q=0.25, and that these values were relatively stable throughout training ($sd$ = 0.015). A paired-samples $t$-test showed that average modularity across subjects was significantly higher for the real networks ($\bar{x}$ =0.18) than for the random networks ($\bar{x}$ =0.04), $t(9)$=9.81, $p$ < 0.001. Panel B shows that the mean number of modules was between 4 and 5, and that these values were relatively stable throughout training ($sd$ = 0.252). The mean number of modules was significantly smaller for the real networks ($\bar{x}$ =4.32) than for the random networks ($\bar{x}$ =7.72), $t(9)$=19.09, $p$ <0 .001. These results confirm previous reports suggesting that functional brain networks have higher modularity than expected by chance (Alexander-Bloch et al., 2012; Bassett et al., 2011; Meunier et al., 2009).

## 5. Head motion analysis

Several recent studies have shown that in-scanner head motion artificially modifies measures of functional connectivity in resting-state fMRI studies (Power et al., 2012, 2014; Satterthwaite et al., 2012; Van Dijk et al., 2012). Such motion confounds can be attenuated, but not completely eliminated through preprocessing steps such as motion correction and confound regression (Satterthwaite et al., 2013). Depending on how the imaging data are pre-analyzed, functional connectivity can increase for short-range connections and decrease for long-range connections (Power et al., 2012, 2014; Satterthwaite et al., 2012; Van Dijk et al., 2012) or it can increase more or less homogenously for all connections (Satterthwaite et al., 2013). The influence of head motion can be reflected in measures derived from network theory, such as those used in the present study. For example, modularity decreases with increasing head motion (Satterthwaite et al., 2012).

The influence of head motion on task-based functional connectivity has not yet been studied, but some authors propose that the artifact must be present in this case as well (Power et al., 2012). If this is correct, then these motion artifacts could be particularly problematic in studies that focus on studying variability on functional connectivity measures across participants and learning stages (Van Dijk et al., 2012), such as the present study. More concretely, if participants in our study moved less as they adapt to the scanner environment, then this could artificially produce stronger functional connectivity in earlier than later sessions. Because performance in the task increases with experience in the task, such motion-based variability in functional connectivity could be correlated with performance. Additionally, participants could show individual differences in average head motion (Van Dijk et al., 2012) that might correlate with motivation to learn the task or other factors explaining individual differences in performance. Such correlations between motion and performance could be higher for accuracy than for response time, or vice-versa, as the correlation between these two behavioral measures is quite small (see Figure 1E in the main article). This could explain some of the dissociations between accuracy and response time found in our study. Results that are specific to particular regions of interest or area types are more difficult to explain as resulting from motion artifacts, because motion affects all functional connectivity between all brain areas more or less uniformly (Satterthwaite et al., 2013). If anything, the confounding effect of motion should be smaller in connections emanating from subcortical regions, which are close to the pivot point for head rotation (Satterthwaite et al., 2013). However, in many of our results such regions played the most important role.

To determine whether any aspects of our results can be explained as resulting from variability in head motion that is correlated with changes in performance, we performed an analysis in two steps. The first step was aimed at determining whether variability in head motion across participants and training blocks could predict behavioral performance, as network measures do.

Motion correction with MCFLIRT provides estimates of three translation and three rotation parameters representing displacement from a reference volume in the middle of the time series (Jenkinson et al., 2002). From these values, we computed the absolute value of displacement of each volume relative to the previous volume. The mean of these displacement values, called the mean relative displacement (MRD; Satterthwaite et al., 2013), was used to summarize head motion in each functional scan. Here we computed the MRD separately from each motion parameter, which resulted in a vector of six MRD values that were used as predictors in two multilevel regression analyses: one with transformed accuracy and one with transformed response times as outcome variable. The model used and the fitting procedures were identical to those used in the main multilevel regression analysis (see Methods in main document). We also fit a null model consisting of only intercept parameters, which was used in a likelihood ratio test to determine whether the motion predictors produced a significant increase in fit to the data.

The results of this analysis indicated that the motion parameters did not significantly increase the model's fit to the data, either when accuracy was the outcome variable, $\chi^2(6) = 8.75$, $p > 0.10$, or when correct response time was the outcome variable, $\chi^2(6) = 4.62$, $p > 0.50$. That is, variability in head motion could not explain changes in performance like network measures do.

The second step of our head-motion analysis was aimed at determining whether the results from our main analysis would still hold after removing the influence of head motion from the behavioral variables. Regressing-out summary statistics of motion in the group-level analysis has been proposed as a way to control for the effects of motion that persist after preprocessing at the individual level (Power et al., 2014; Satterthwaite et al., 2012).

We used the previously-fitted models (residuals and intercept parameters) to compute the values of behavioral variables (accuracy and response time) after the influence of motion was regressed out. All steps in the main regression analysis (see main manuscript) were repeated using these motion-corrected outcome variables. The results are shown in Figures S3-S8 and the results of statistical tests are summarized in the legend of each figure ($\alpha$=0.05 for all tests). Most results from the main analysis are reproduced in the analysis of motion-controlled data: the model selection procedure resulted in most best-fit models being simple (Figure S3A), subcortical areas were very important for the explanation of accuracy and not for the explanation of response time (Figure S3B), the direction of biases in regression coefficient distributions were identical to those reported in the main manuscript (Figures S4-S8), visual areas had mean regression coefficients close to zero for several predictors (Figures S5-S7), and subcortical areas had high-magnitude mean regression coefficients for several predictors (Figure S4B-right, Figure S6B-left and Figure S7B-left). In all of these cases, the effects found here were in the same direction and comparable to those reported in the main analysis, although some of them were not significant. Our conclusions and discussion in the main manuscript focus on the aspects of our results that were robust to motion correction.

In sum, most differences between the results of the present analysis and those of the main analysis were not qualitative, but quantitative, affecting the significance of some effects but not their direction. This was to be expected, for two reasons. First, because eliminating the effect of motion on the outcome variables through regression can eliminate variability that is important for the main effect under study (Power et al., 2014; Satterthwaite et al., 2012), which in this case was variability in performance that co-varies with variability not due to motion in the network measures. This problem is likely to reduce the strength of effects in all statistical analyses. Second, the number of nodes classified as predictive was lower in this analysis than in the main analysis (see Figure S3), resulting in a reduction in power of statistical tests for the analysis of regression coefficients. Part of the reason for a lower number of predictive nodes was that variance of interest was regressed out from the behavioral variables.

Surprisingly, in some cases effects that were numerically present in the main analysis, but not significant, were found to be significant in the present analysis. In the analysis of the participation coefficient (Figure S6), the low mean ranking of motor areas in the response time analysis was significant only after motion correction. In the analysis of intra-modular degree (Figure S7), mean ranking of motor areas in the accuracy analysis was high, whereas mean ranking of high-level areas in the response time analysis was low, and in both cases the effect was significant only after motion correction. In the analysis of betweenness centrality (Figure S8), the low mean ranking of motor areas in the accuracy analysis was significant only after motion correction.

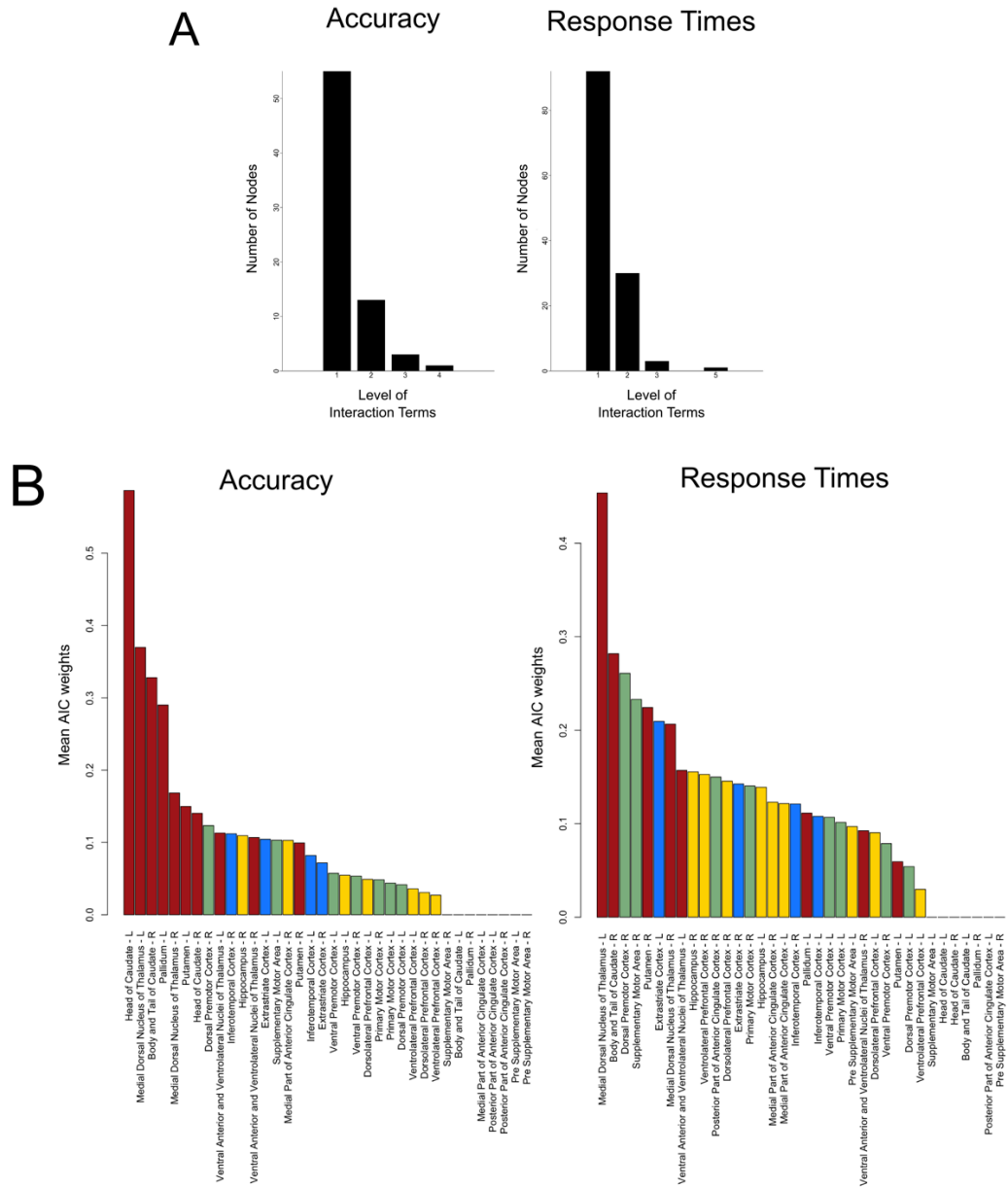

**Figure S3.** Characterization of best-fitting models in predictive nodes in terms of their distribution of complexity and fit to the data, after motion correction of behavioral data through regression. Compare to Figure 2 of the main manuscript. (A) Distribution of the level of interaction terms for the best-fitting model across predictive nodes; although the number of nodes classified as predictive is smaller than in the main analysis (72 for accuracy and 126 for response time), their distribution is still characterized by a majority of nodes without interaction terms. (B) Distribution of AIC weights for the best-fitting model of predictive nodes across regions of interest. Mean ranking of subcortical areas was higher than chance for accuracy and significantly different in both distributions. Mean ranking of high-order areas was lower than chance for accuracy. No other tests were significant.

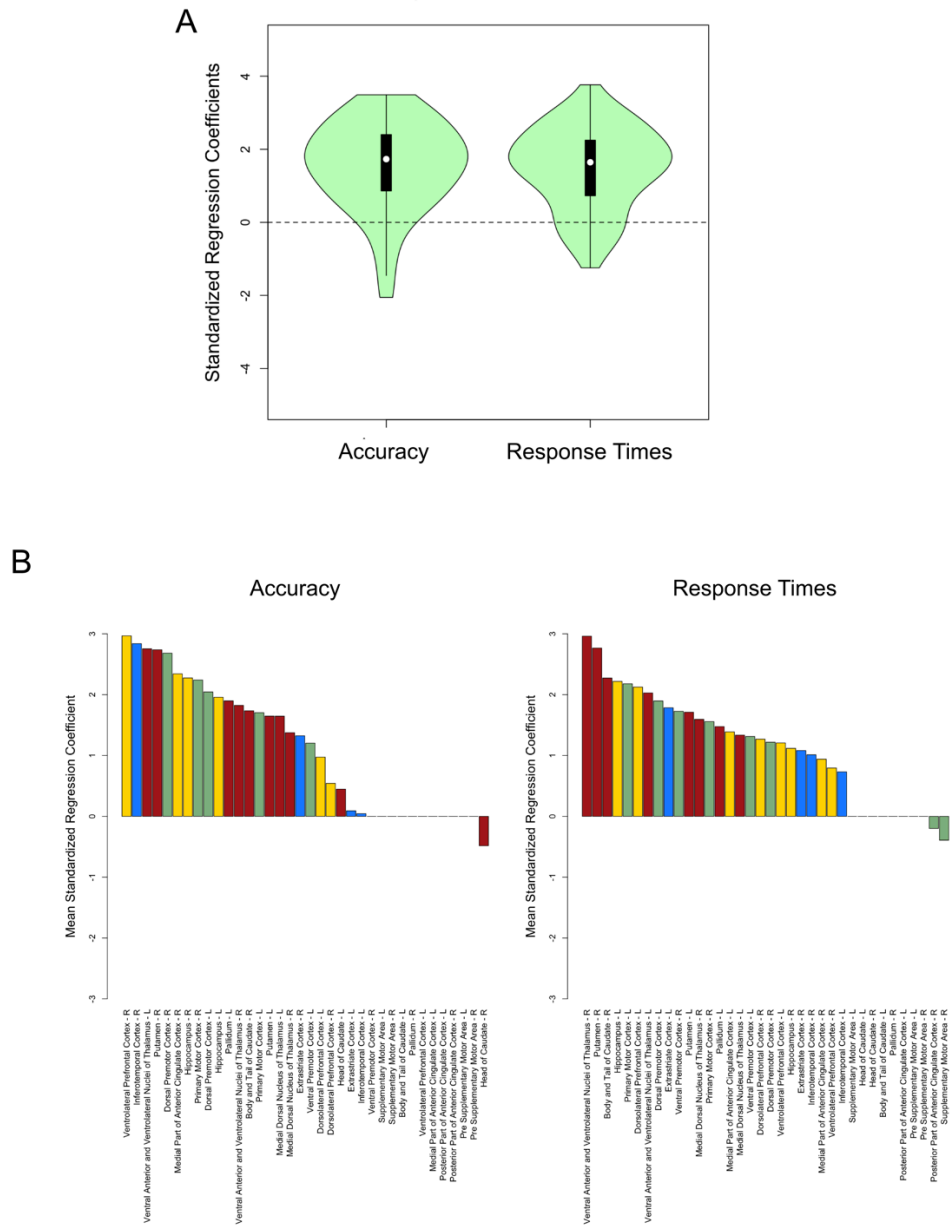

**Figure S4.** Distribution of standardized regression coefficients for clustering coefficient, after motion correction of behavioral data through regression. Compare to Figure 3 of the main manuscript. (A) Distribution across all predictive nodes; both distributions show a significant positive bias, but they do not differ from each other. (B) Mean standardized regression coefficient across regions of interest; none of the permutation tests was significant.

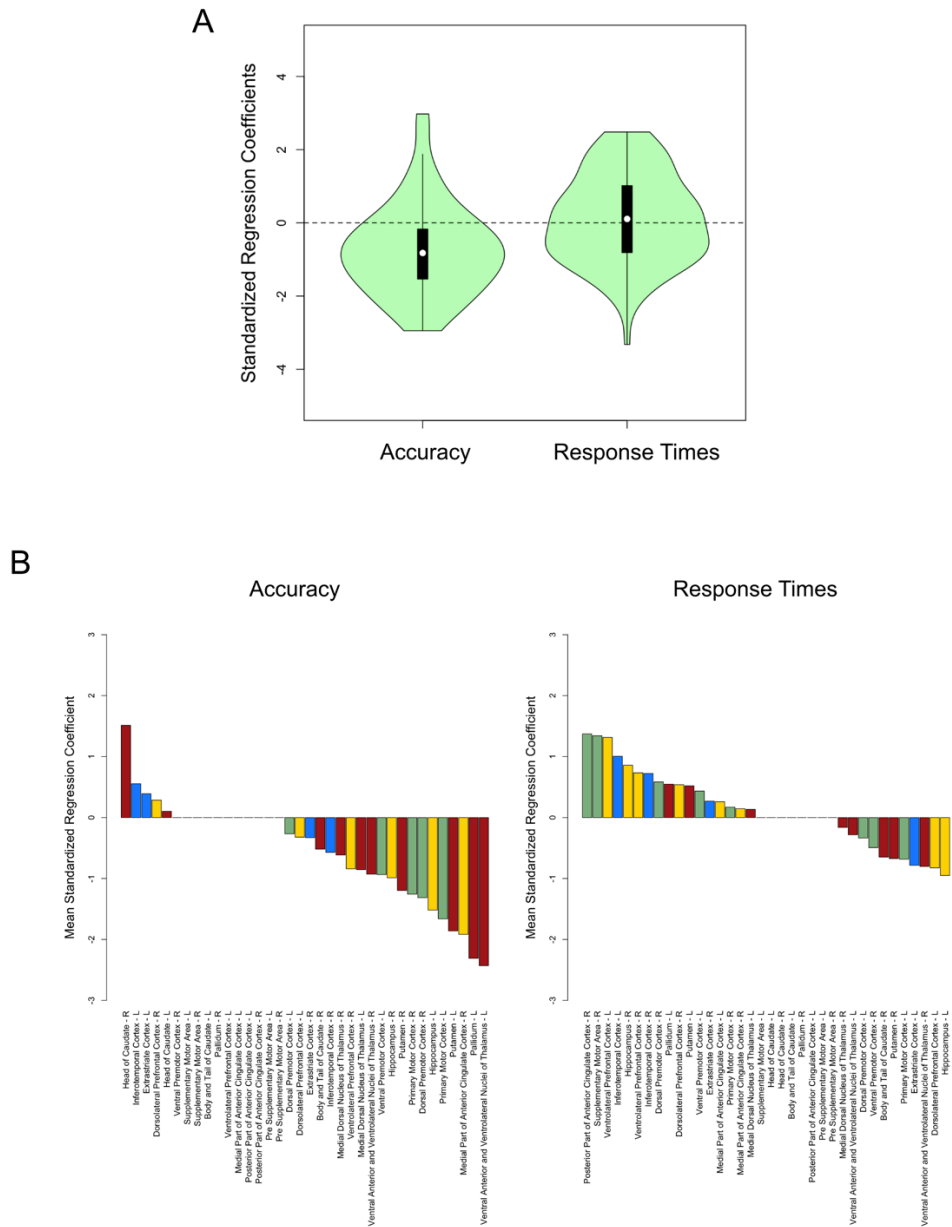

**Figure S5.** Distribution of standardized regression coefficients for strength, after motion correction of behavioral data through regression. Compare to Figure 4 of the main manuscript. (A) Distribution across all predictive nodes; the distribution for accuracy shows a significant negative bias; no other tests are significant. (B) Mean standardized regression coefficient across regions of interest. Mean ranking of visual areas was higher than chance in the accuracy distribution; no other permutation test was significant.

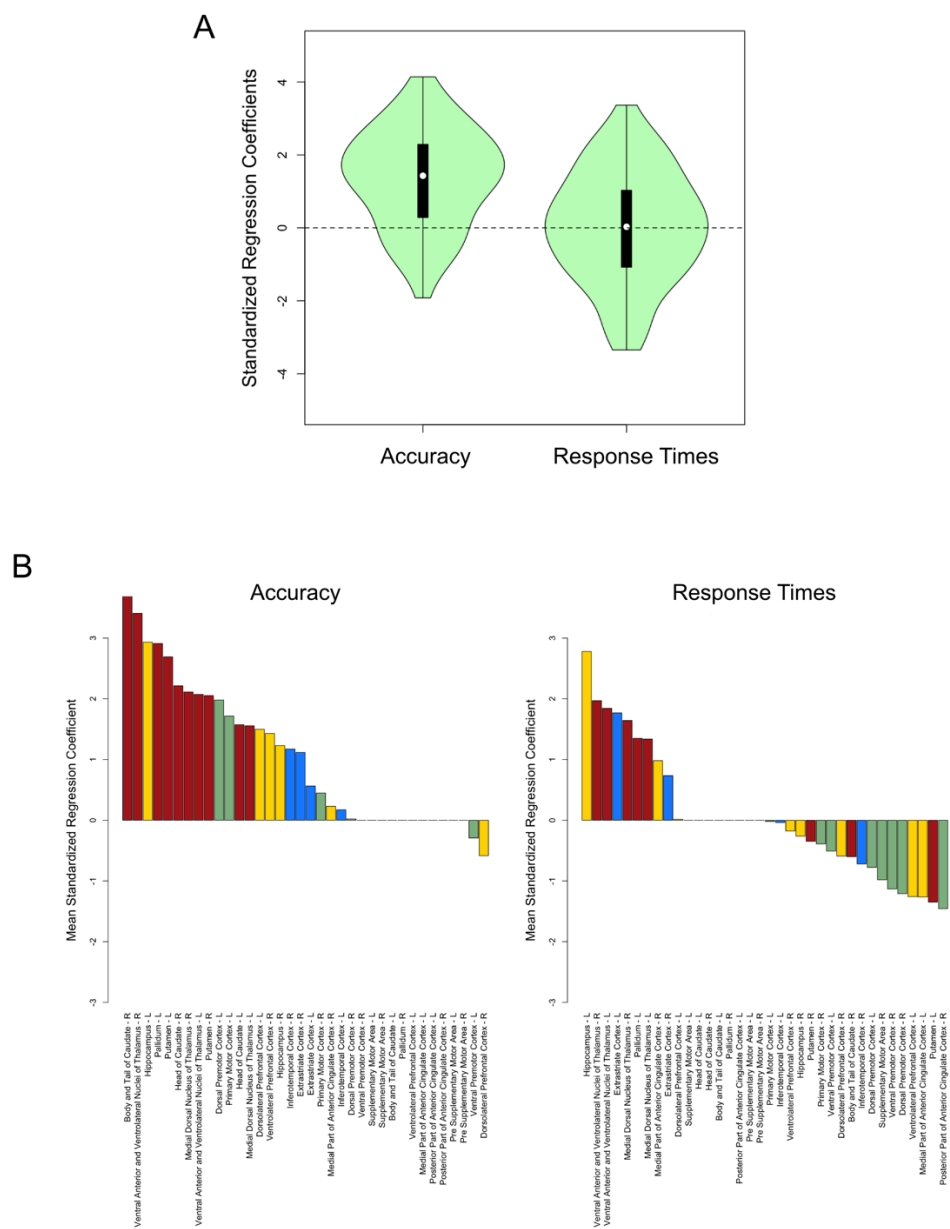

**Figure S6.** Distribution of standardized regression coefficients for participation coefficient, after motion correction of behavioral data through regression. Compare to Figure 5 of the main manuscript. (A) Distribution across all predictive nodes; the distribution for accuracy shows a significant positive bias; no other tests are significant. (B) Mean standardized regression coefficient across regions of interest. In the accuracy distribution, mean ranking was higher than chance in subcortical areas and lower than chance in visual areas. Mean ranking was higher than chance for motor areas in the response time distribution. The difference in mean ranking of visual areas was significant. No other tests were significant.

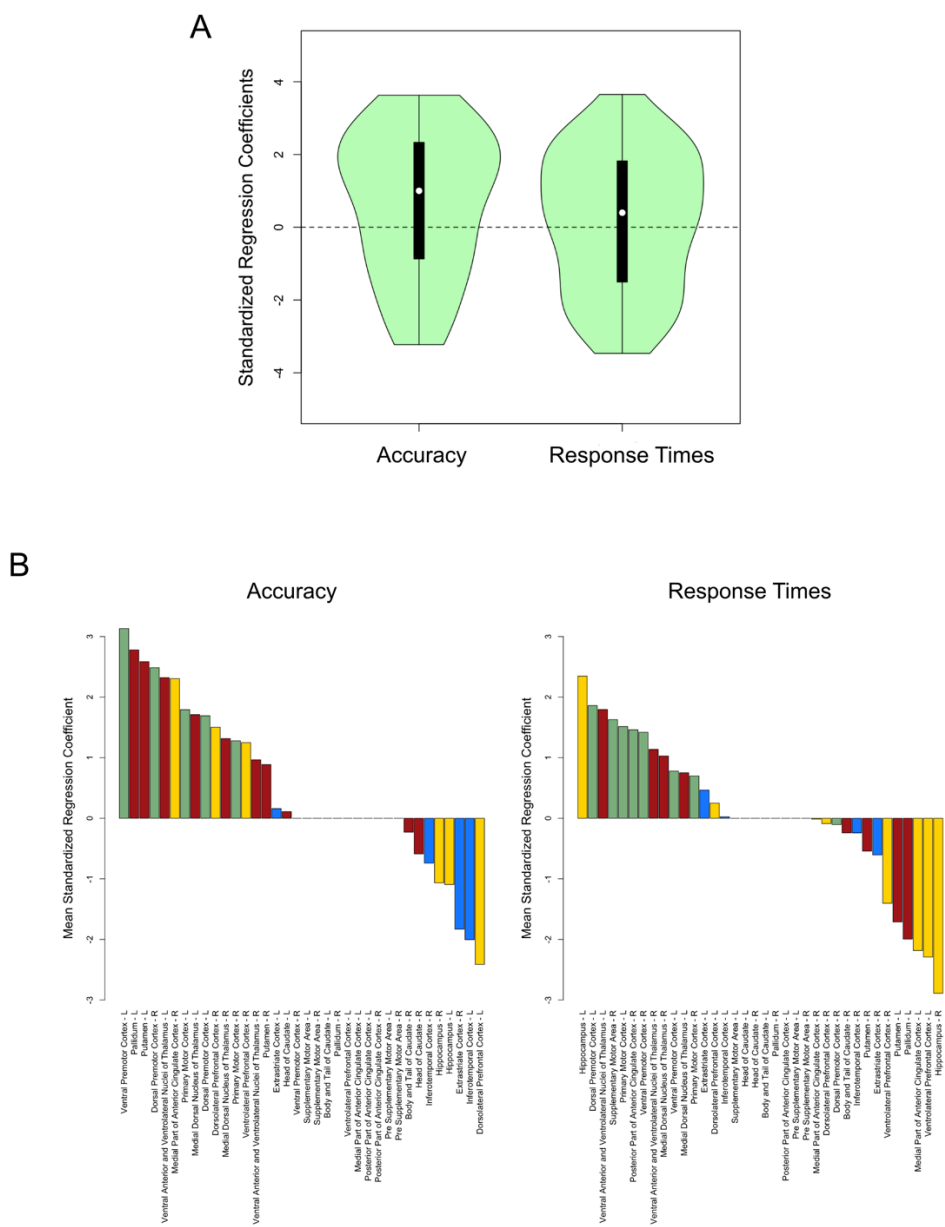

**Figure S7.** Distribution of standardized regression coefficients for intra-modular strength, after motion correction of behavioral data through regression. Compare to Figure 6 of the main manuscript. (A) Distribution across all predictive nodes; no tests were significant. (B) Mean standardized regression coefficient across regions of interest. In the accuracy distribution, mean ranking was higher than chance in motor areas and lower than chance in visual areas. Mean ranking was higher than chance for high-level areas in the response time distribution. No other tests were significant.

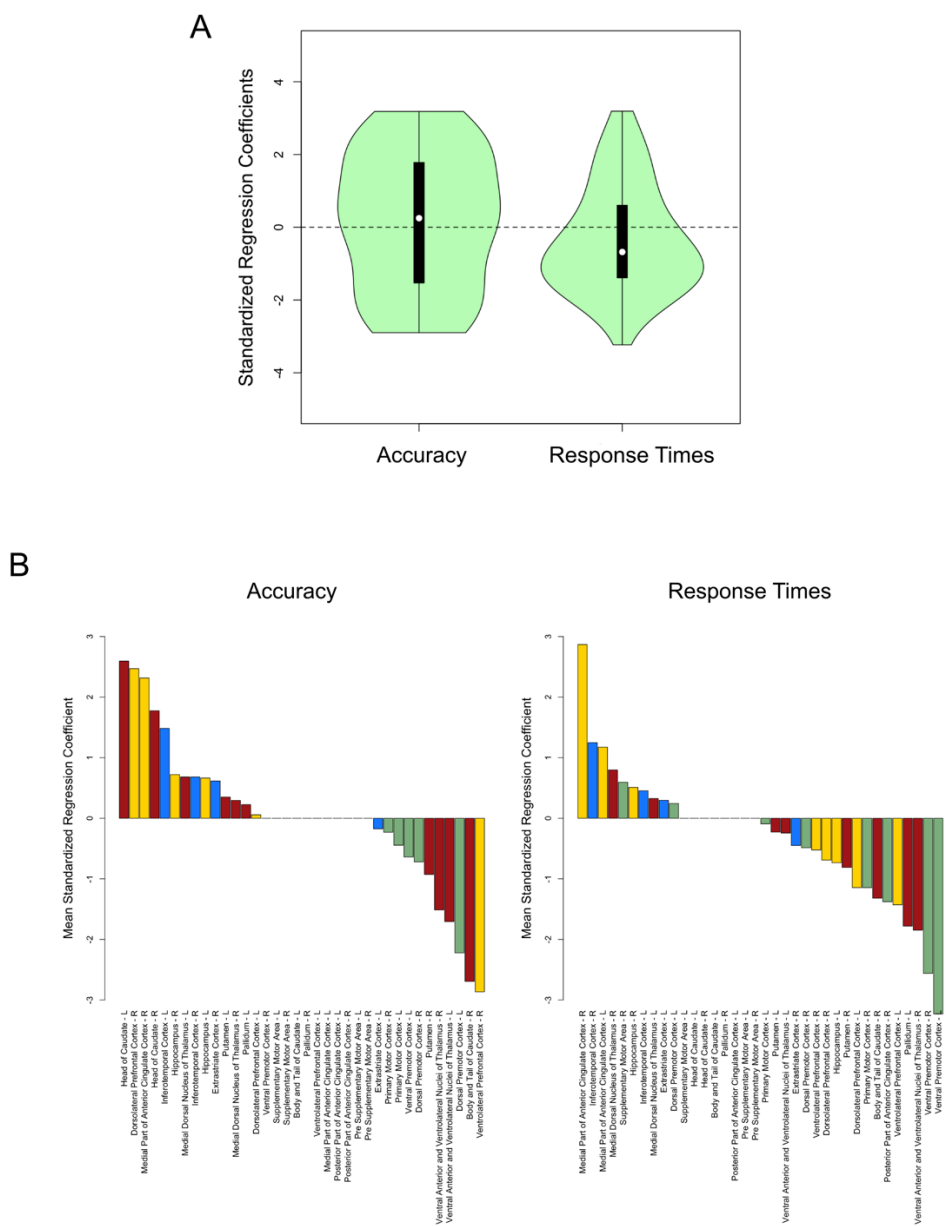

**Figure S8.** Distribution of standardized regression coefficients for betweenness centrality, after motion correction of behavioral data through regression. Compare to Figure 7 of the main manuscript. (A) Distribution across all predictive nodes; the distribution for response time shows a significant negative bias; no other tests are significant. (B) Mean standardized regression coefficient across regions of interest. Mean ranking was higher than chance for motor areas in the accuracy distribution. No other tests were significant.

# References

Alexander-Bloch A, Lambiotte R, Roberts B, Giedd J, Gogtay N, Bullmore E (2012) The discovery of population differences in network community structure: New methods and applications to brain functional networks in schizophrenia. NeuroImage 59:3889–3900.

Bassett DS, Wymbs NF, Porter MA, Mucha PJ, Carlson JM, Grafton ST (2011) Dynamic reconfiguration of human brain networks during learning. Proc Natl Acad Sci 108:7641-7646.

Bell BA, Morgan GB, Schoeneberger JA, Kromrey JD, Ferron JM (2014) How low can you go? An investigation of the influence of sample size and model complexity on point and interval estimates in two-level linear models. Methodol Eur J Res Methods Behav Soc Sci 10:1–11.

Benjamini Y, Hochberg Y (1995) Controlling the false discovery rate: A practical and powerful approach to multiple testing. J R Stat Soc Ser B Methodol 57:289–300.

Dormann CF, Elith J, Bacher S, Buchmann C, Carl G, Carré G, Marquéz JRG, Gruber B, Lafourcade B, Leitão PJ (2013) Collinearity: A review of methods to deal with it and a simulation study evaluating their performance. Ecography 36:027–046.

Gelman A, Hill J (2007) Data Analysis Using Regression And Multilevel/Hierarchical Models. New York, NY: Cambridge University Press.

Jenkinson M, Bannister P, Brady M, Smith S (2002) Improved optimization for the robust and accurate linear registration and motion correction of brain images. NeuroImage 17:825–841.

Maas CJM, Hox JJ (2004) The influence of violations of assumptions on multilevel parameter estimates and their standard errors. Comput Stat Data Anal 46:427–440.

Maas CJM, Hox JJ (2005) Sufficient sample sizes for multilevel modeling. Methodol Eur J Res Methods Behav Soc Sci 1:86.

Maslov S, Sneppen K (2002) Specificity and stability in topology of protein networks. Science 296:910–913.

Meunier D, Achard S, Morcom A, Bullmore E (2009) Age-related changes in modular organization of human brain functional networks. NeuroImage 44:715–723.

Murdoch DJ, Chow ED (1996) A graphical display of large correlation matrices. Am Stat 50:178–180.

Power JD, Barnes KA, Snyder AZ, Schlaggar BL, Petersen SE (2012) Spurious but systematic correlations in functional connectivity MRI networks arise from subject motion. NeuroImage 59:2142–2154.

Power JD, Mitra A, Laumann TO, Snyder AZ, Schlaggar BL, Petersen SE (2014) Methods to detect, characterize, and remove motion artifact in resting state fMRI. NeuroImage 84:320–341.

Rawlings JO, Pantula SG, Dickey DA (1998) Applied Regression Analysis: A Research Tool. New York, NY: Springer.

Rubinov M, Sporns O (2010) Complex network measures of brain connectivity: Uses and interpretations. NeuroImage 52:1059–1069.

Satterthwaite TD, Elliott MA, Gerraty RT, Ruparel K, Loughead J, Calkins ME, Eickhoff SB, Hakonarson H, Gur RC, Gur RE, Wolf DH (2013) An improved framework for confound regression and filtering for control of motion artifact in the preprocessing of resting-state functional connectivity data. NeuroImage 64:240–256.

Satterthwaite TD, Wolf DH, Loughead J, Ruparel K, Elliott MA, Hakonarson H, Gur RC, Gur RE (2012) Impact of in-scanner head motion on multiple measures of functional connectivity: Relevance for studies of neurodevelopment in youth. NeuroImage 60:623–632.

Shapiro SS, Wilk MB (1965) An analysis of variance test for normality (complete samples). Biometrika 52:591–611.

Snijders TAB, Bosker RJ (2012) Multilevel Analysis: An Introduction to Basic and Advanced Multilevel Modeling, 2nd ed. Thousand Oaks, CA: SAGE.

Van Dijk KRA, Sabuncu MR, Buckner RL (2012) The influence of head motion on intrinsic functional connectivity MRI. NeuroImage 59:431–438.